\documentclass[10pt, aps,prl,twocolumn,floatfix,superscriptaddress]{revtex4-2}

\usepackage{graphicx}
\usepackage{amsmath}
\usepackage{amssymb}
\usepackage{hyperref}
\usepackage{xcolor}
\usepackage{dsfont}
\usepackage{enumitem}

\begin{document}
\UseRawInputEncoding

\title{Multicomponent Kardar-Parisi-Zhang Universality in Degenerate Coupled Condensates}

\author{H.~Weinberger}
\email[Corresponding author: ]{harvey.weinberger.22@ucl.ac.uk} 
\affiliation{Department of Physics and Astronomy, University College London, 
Gower Street, London, WC1E 6BT, United Kingdom}

\author{P.~Comaron}
\affiliation{Department of Physics and Astronomy, University College London, 
Gower Street, London, WC1E 6BT, United Kingdom}
\affiliation{CNR NANOTEC, Institute of Nanotechnology, Via Monteroni,  73100 Lecce, Italy}

\author{M.H.~Szyma\'nska}
\affiliation{Department of Physics and Astronomy, University College London, 
Gower Street, London, WC1E 6BT, United Kingdom}

\begin{abstract}
We show that the multicomponent Kardar-Parisi-Zhang equation describes the low-energy theory for phase fluctuations in a \(\mathbb{Z}_{2}\) degenerate non-equilibrium driven-dissipative condensate with global \(U(1)\times U(1)\) symmetry. Using dynamical renormalisation group in spatial dimension \(d=1\), we demonstrate that coupled stochastic complex Ginsburg-Landau equations exhibit an emergent stationary distribution, enforcing KPZ dynamical exponent \(z=3/2\) and static roughness exponent \(\chi=1/2\) for both components. By tuning intercomponent interactions, the system can access other regimes, including a fragmented condensate regime from a dynamical instability in the phase fluctuations, as well as a spacetime vortex regime driven by the non-linear terms in the coupled KPZ equations. In stable regimes, we show that in specific submanifolds relevant to polaritons, the RG fixed point offers a transformation to decoupled KPZ equations. 
Our findings have broad implications for understanding multicomponent KPZ systems in the long-wavelength limit. 
\end{abstract}

\maketitle
The Kardar-Parisi-Zhang (KPZ) universality class has become a prototypical model for non-equilibrium critical phenomena owing to the ever-increasing list of systems it describes. Initially proposed to model rough geometries in stochastic growth processes \cite{PhysRevLett.56.889}, it is notable for its distinct anomalous diffusion, characterised by a dynamical exponent \(z=3/2\) and roughness exponent \(\chi=1/2\) in spatial dimension \(d=1\). As a classical stochastic equation, it has strikingly been observed in quantum systems, particularly as a semiclassical description of phase fluctuations in driven-dissipative condensates such as exciton-polaritons \cite{PhysRevX.5.011017}, and even in transport properties of the Heisenberg spin chain \cite{doi:10.1126/science.abk2397, Scheie2021}. In the single component case, the KPZ equation is largely considered solved, with its statistical properties constrained by Galilean and tilt-symmetries directly enforcing \(\chi+z=2\) and its stationary two-point correlators fully determined \cite{doi:10.1142/S2010326311300014, Pr_hofer_2004}.

A natural question to pose is whether KPZ systems \cite{PhysRevLett.56.889, PhysRevLett.79.1515, HALPINHEALY1995215, Derrida1992, Blythe_2007,Popkov2003, Prähofer2002, PhysRevB.91.045301, PhysRevX.5.011017,PhysRevB.92.155307,Fontaine_2022} offer multicomponent generalisations \cite{PhysRevLett.69.929,10.1016/j.jfa.2017.05.002}. This question has arisen in various cases: dynamic roughening of directed lines \cite{PhysRevLett.69.929}; sedimenting colloidal suspensions \cite{Levine1998}; stochastic lattice gases \cite{Ferrari_2013,PhysRevLett.112.200602,Das2001}; magnetohydrodynamics \cite{Yanase1997NewOM,Fleischer1998}; dynamics of fluids and quantum fluids \cite{vanBeijeren2012}; dynamics of anharmonic chains on a mesoscopic scale \cite{Mendl2013,Spohn2014}; and proliferating active matter~\cite{Hallatschek2023}. It has also been argued that multicomponent KPZ-like equations arise from continuity equations in the non-linear fluctuating hydrodynamics of quantum integrable systems \cite{Ferrari_2013, Roy_2024}. 
The isotropic spin chain at finite temperature is an intriguing instance of this, where transport properties are modelled by coupled Burgers' equations \cite{PhysRevLett.131.197102}. Multicomponent KPZ is directly relevant to polariton systems \cite{pmid17006506,Lagoudakis2008, PhysRevLett.101.067404}, which have established themselves as key platforms for realising non-equilibrium critical behaviour~\cite{PhysRevLett.121.095302,kai2015,alnatah2024critical}. Polaritons are hybridised light-matter quasiparticles typically comprising two optically active spin components \(J_{z}\in \{-1,1\}\) \cite{PhysRevLett.69.3314, 2010SeScT..25a3001S} and a photonic component \cite{PhysRevLett.92.017401}. In the absence of external magnetic fields (an in-plane magnetic field can be generated for polaritons via TE-TM splitting in the sample \cite{PhysRevB.81.045318}), condensates support linear polarisation with equal occupations of spin-up and spin-down states \cite{KLOPOTOWSKI2006511, Shelykh_2010,PhysRevB.93.115313, PhysRevA.89.033631}. Most models ignore this spinor structure. However, when including both fields, it is not \textit{a priori} true that: each mode's scaling remains within the KPZ universality class, that a stationary measure exists, or that the dynamics are constrained by Galilean and tilt-symmetries. This means that systems described by multicomponent KPZ do not generally fit within the single component framework, thus demanding thorough classification while offering an opportunity to realise novel phases in non-equilibrium~\cite{galaa2comps,PhysRevResearch.5.043286}. 

In this letter, we explore the rich physics of multicomponent KPZ, specifically focusing on how it describes the dynamics of the gapless Nambu-Goldstone (NG) modes in driven bosonic systems at weak noise. We fully determine the phase diagram of the linearly-polarised degenerate condensate model with \(\mathbb{Z}_{2}\) inversion symmetry \(\psi_{1} \leftrightarrow \psi_{2}\) highlighting its miscible-immiscible transition. We identify parameter regimes where the effective KPZ equations are unstable, giving rise to a non-thermal vortex turbulent phase. At the physical level, this regime exhibits a dramatically modified exponent \(z=1\) due to phase compactness. We construct the most general \(\mathbb{R}^{2}\)-KPZ equations obeying \(\mathbb{Z}_{2}\) internal symmetries for the unwound phase and probe the infrared physics to one-loop in Wilsonian renormalisation group (RG) using MSRJD formalism \cite{PhysRevB.67.115412,PhysRevLett.128.070401, PhysRevE.52.4741, täuber_2014, honkonen2012ito}.  
This maps the stochastic equation into a \(\mathbb{R}^{d+1}\) classical field theory for the phase modes \((\theta^{\alpha})_{\alpha=1}^{2}\) amenable to perturbative treatment around their Gaussian free theories. In both RG analysis and direct integration of the multicomponent KPZ, we recover the critical exponents \(z=3/2\) and \(\chi=1/2\) in stable regimes. Finally, we highlight non-universal effects on the distribution of fluctuations within specific submanifolds, such as the Cole-Hopf line where a non-trivial decoupling transformation exists. This analysis provides a comprehensive understanding of the long-wavelength behaviour of coupled condensates, but applies to many other systems.  
\begin{figure*}[ht!]
    \centering
    \vspace{-0.5cm}
    \hspace{-1cm}
    \includegraphics[width=0.9\textwidth]{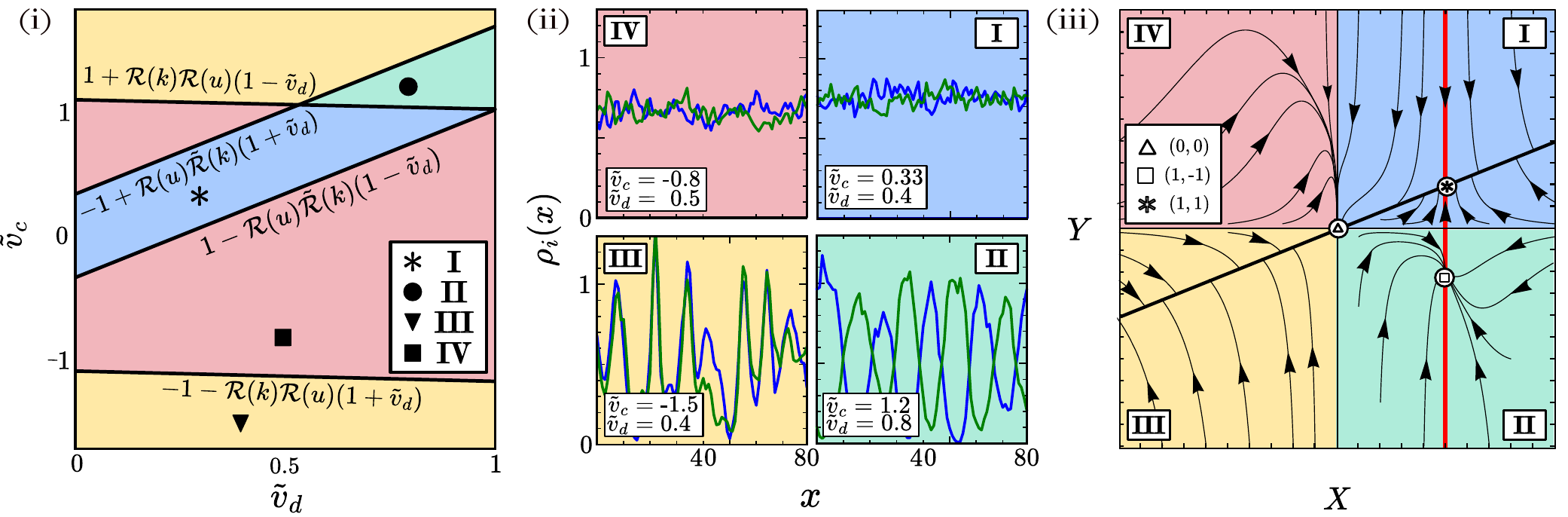}
    \begin{minipage}{\textwidth}
        \vspace{-0.25cm}
        \caption{(i) Stability conditions in \((\tilde{v}_{d},\tilde{v}_{c})\) for Eq.~\eqref{eq:coupled_SCGLE} with \(\mu_{d}=u_{d}=k_{d}=1, \,u_{c}=3,\, k_{c}=4\) and \(\tilde{\mathcal{R}}(k)=k_{c}/k_{d}\); hence \(t\) and \(x\) are measured in \(\mu_{d}^{-1}\) and \(\sqrt{k_{d}}\). Tuning \(\tilde{v}_{c}\) and \(\tilde{v}_{d}\) allows crossing between quadrants in (iii), denoted by the same colours across all subfigures signifying different regimes. (ii) Density fluctuations for \(\theta_{1},\, \theta_{2}\) (blue/green) components for parameters in Quadrants (I-IV). (iii) Projected RG flows in the \((X,Y)\) plane for \( \tilde{D}_{22}/\tilde{D}_{11}=1\) with Cole-Hopf \(X=1\) (red) and FDR \(X=Y\) (black) lines. For the SCGLE, Quadrant I gives KPZ scaling in both modes, Quadrant IV is an STV-dominated phase, while Quadrants II and III have a dynamical instability in the gapless modes, making the mapping inappropriate.}
     \label{fig:combined}
    \end{minipage}
\end{figure*}
\paragraph{Degenerate Coupled Condensates} - Typically, mesoscopic dynamics of driven-dissipative condensates in \(d=1\) are modelled by coupled SCGLE for the lower polariton branch fields \(\psi_{i} \in \mathbb{C}\) \cite{He_2017}
\begin{equation}
\begin{aligned}
    i \partial_{t} \psi_{i} &= \Big[-(k_{c}-ik_{d})\partial_{x}^{2}+(\mu_{c}+i\mu_{d})+\\&(u_{c}-iu_{d})|\psi_{i}|^{2}+(v_{c}-iv_{d})|\psi_{\overline{i}}|^{2}\Big]\psi_{i} + \zeta_{i},
\end{aligned}
\label{eq:coupled_SCGLE}
\end{equation}
with a density coupling to the other component \(\psi_{\overline{i}}\). This model can be derived from Keldysh field theory in the semiclassical approximation assuming a Markovian coupling to the decay and pumping bath modes. All parameters are real and \(\mu_{d},\,u_{d},\, u_{c}>0\) for a stable condensed solution, and \(\zeta_{i}\) is a Gaussian complex white noise with diagonal correlations \(\langle\zeta^{*}_{i}(x,t)\zeta_{j}(x',t)\rangle=(\gamma_{p}+\gamma_{l})\delta_{ij}\delta(x-x')\delta(t-t')\). The kinetic coefficient is \(k_{c}=1/2m_{LP}\) where \(m_{LP}\) is the effective polariton mass and \(k_{d}\) is a diffusion constant which acts to suppress higher momentum modes. The system is driven-dissipative requiring incoherent external pumping from a laser, and is in contact with an excitonic decay bath. This gives rise to an effective gain \(\mu_{d}=(\gamma_{p}-\gamma_{l})/2\) from the incoherent pump minus the single particle losses which is positive in the \(U(1)\) spontaneously symmetry broken (SSB) condensate phase. The two particle losses \(u_{d},\,v_{d}>0\) act as gain saturation terms. This model assumes local interactions which is valid provided we ignore the spatial extent of the excitons. For polaritons, the intercomponent, \(v_{c}\), and intracomponent, \(u_{c}\), interactions are attractive and repulsive respectively. Typically, experimental cross-spin scattering rates are \(5-10\%\) of the intraparticle scattering rate implying \(v_{c} \approx -0.1 u_{c}\) \cite{PhysRevB.75.045326,PhysRevB.81.235302}.

Equilibrium two-component condensates support a miscible-immiscible transition when the real intercomponent interactions exceed the intracomponent interactions \(v_{c}^{2}>u_{c}^{2}\) \cite{PhysRevLett.78.586,PhysRevLett.81.1539,PhysRevLett.107.193001,PhysRevLett.81.5718} where modulation instabilities cause condensate fragmentation. Non-equilibrium condensates support a similar phase separation \cite{PhysRevB.92.155307, PhysRevA.89.033631} but differ from their equilibrium counterparts in that the linearised SCGLE \eqref{eq:coupled_SCGLE} no longer has four gapless sound modes in its Bogoliubov spectrum but instead has two gapped density modes and two gapless diffusive phase modes. There are two conditions of different physical origin which can drive a miscible transition shown in FIG.~\ref{fig:combined}(i-ii): (a) the dissipative intercomponent interaction exceeds the intracomponent interactions \(v_{d}>u_{d}\), resulting in the gapped modes becoming dynamically unstable, and (b) when the rescaled intercomponent interaction \(\tilde{v}_{c}=v_{c}/u_{c}\) does \emph{not} satisfy
\begin{equation}
    -1 - \mathcal{R}(k)\mathcal{R}(u)(1+\tilde{v}_{d}) <\tilde{v}_{c} < 1 + \mathcal{R}(k)\mathcal{R}(u)(1-\tilde{v}_{d})
\label{eq:conditionsstability}
\end{equation}
with \(\mathcal{R}(k)=k_{d}/k_{c}\), resulting in the gapless modes becoming dynamically unstable. Within the \(\tilde{v}_{d}= v_{d}/u_{d}<1\) region, there exist two sub-cases for the miscible transition depending on whether the intercomponent interaction \(v_{c}\) is attractive or repulsive. For repulsive intercomponent interactions \(v_{c}>0\), tuning \((\tilde{v}_{c},\tilde{v}_{d})\) outside of the immiscible boundary Eq.~\eqref{eq:conditionsstability}, gives rise to spatially localised single component regions. For these parameters which map to Quadrants II and III in FIG.~\ref{fig:combined}(i-iii), the low-energy sector is described by single component KPZ if density fluctuations are sufficiently irrelevant \cite{He_2017}. If the interaction is attractive \(\tilde{v}_{c}<0\), as shown for the density plot for the simulation in Quadrant III in FIG.~\ref{fig:combined}(ii), there is no obvious fragmentation. However, the density fluctuations are still large and cannot be adiabatically eliminated, making the KPZ mapping inappropriate.

The theory is invariant under \(U(1)\times U(1)\) symmetry with independent \(\psi_{i} \mapsto e^{i\theta_{i}}\psi_{i}\). When \(\mu_{d}>0\), each \(U(1)\) undergoes SSB. The ground state topology dictates that the effective theory is parameterised using the density-phase representation where \(\theta_{i}\) parameterise the gapless fluctuations. In the adiabatic approximation around the mean field \(|\psi_{0,i}|^{2}= \rho_{i} = {\mu_{d}}/{(u_{d}+v_{d})} = - {\mu_{c}}/{(u_{c}+v_{c})}\), gapped fluctuations can be integrated out \cite{Sieberer_2016} giving multicomponent KPZ equations for \(\mathbb{R}^{2}\)-fields \((\theta^{\alpha}(x,t))_{\alpha=1}^{2}\)
\begin{equation}
\begin{aligned}
    \partial_{t} \theta^{\alpha} = D_{\alpha\beta} \partial_{x}^{2} \theta^{\beta} + \dfrac{1}{2}\Gamma^{\alpha}_{\beta\gamma} (\partial_{x} \theta^{\beta})(\partial_{x} \theta^{\gamma}) + \zeta^{\alpha}
\end{aligned}
\label{equation:mckpzequation}
\end{equation}
driven by a real Gaussian additive white noise \((\zeta^{\alpha})_{\alpha=1}^{2}\) with symmetric covariance \(
\langle \zeta^{\alpha}(x,t) \zeta^{\beta}(x',t') \rangle = 2 \Delta^{\alpha \beta} \delta(x-x') \delta(t-t')\). Strictly, the phase itself is a compact variable, supporting non-trivial topological excitations such as spacetime vortices (STV) which modify the scaling. For the single-component case in the low-noise regime, compactness does not affect KPZ scaling for significant time windows \cite{PhysRevLett.118.085301}. The parameters from the SCGLE mapping are
$D_{11} = k_{c}\mathcal{C}_{1}(u_{d}u_{c} - v_{c}v_{d}) + k_{d},\,  
    D_{12} = k_{c}\mathcal{C}_{1}(v_{c}u_{d} - u_{c}v_{d}),\, 
    \Gamma^{1}_{11} = -2 k_{c} + 2k_{d}\mathcal{C}_{1}\left(u_{c}u_{d}-v_{c}v_{d}\right),\,
    \Gamma^{1}_{22} = 2k_{d}\mathcal{C}_{1}(v_{c}u_{d} - u_{c}v_{d}),\, 
    \Gamma^{1}_{12}= 0$,
where \(\mathcal{C}_{1} = (u_{d}^{2} - v_{d}^{2})^{-1}\). The remaining parameters can be inferred from \(\mathbb{Z}_{2}\) symmetry. Coupling in the real density channel between the condensates \(v_{c}\) produces non-trivial couplings in the off-diagonals of the diffusion and noise matrices. The non-linearity vanishes at the equilibrium condition when \(k_{d}/k_{c}=v_{d}/v_{c}=u_{d}/u_{c}\) define collinear rays in the complex plane \cite{PhysRevLett.110.195301}. A point to stress is that off-diagonal interaction vertices \(\Gamma^{\alpha}_{\beta\gamma}\) are only generated if \(k_{d}\neq 0\). The \(\Delta^{\alpha\beta}\) have non-trivial cross-components resulting in correlated white noise driving the two phase modes and are presented in the supplemental materials Ref.~\cite{supp}. Imposing that the diffusion matrix \(D\) is positive definite is equivalent to the dynamical stability condition in Eq.~\eqref{eq:conditionsstability}. 

\begin{figure}[htbp]
    \centering
    \includegraphics[width=0.482\textwidth]{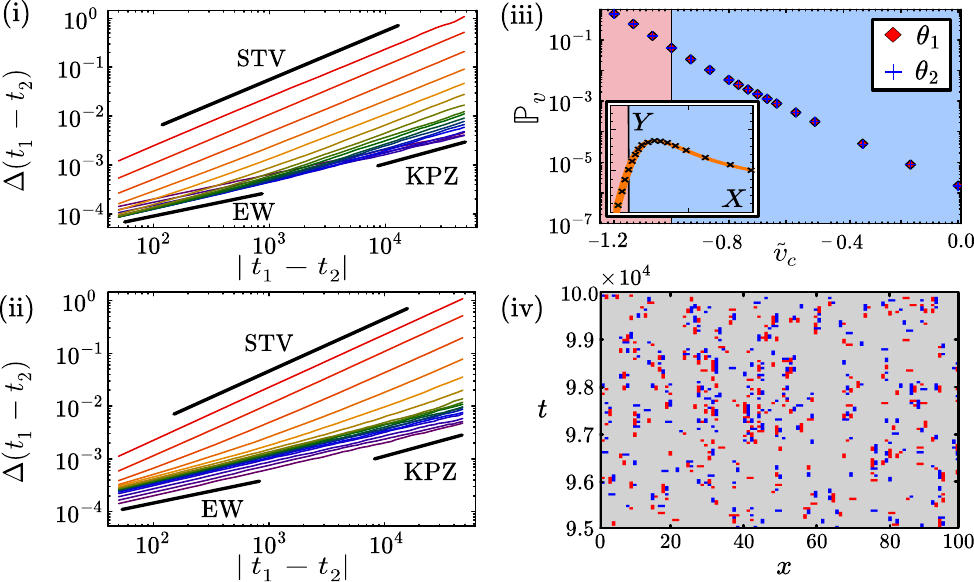}
    \vspace{-0.25cm}
    \caption{SCGLE simulations with parameters \(v_{d}=0\), \(k_{d}=u_{d}=\mu_{d}=1\) and \(k_{c}=3,\;u_{c}=1.5\) and tuning \(\tilde{v}_{c}\) from \(0\) to \(-1.2\). (i-ii) The autocorrelation  \(\Delta(t_{1}-t_{2})\) of the unwound phase for \(\tilde{\theta}_{1}\) and \(\tilde{\theta}_{2}\) respectively showing a crossover to the STV regime when tuning \(\tilde{v}_{c}\) between \(0\) to \(-1.2\) plotted from bottom to top.  (iii) Probability of a vortex in the spacetime lattice for unrotated \(\theta_{1}\), \(\theta_{2}\) as a function of attractive \(\tilde{v}_{c}\). Inset shows the line traced out in the \((X,Y)\) plane in FIG.~\ref{fig:combined}(iii) from the decoupled point \((1,1)\) in Quadrant I (KPZ) in blue to Quadrant IV (STV) in red. (iv) Vortex charge distribution with \(+1\) (red) and \(-1\) (blue) on a section of the spacetime lattice for \(\tilde{v}_{c}=-1.2\) in the STV dominated Quadrant IV.}
    \label{fig:figurestv}
    \vspace{-0.25cm}
\end{figure}

In the spirit of Landau, the effective theory in the multicomponent regimes should be described by Eq.~\eqref{equation:mckpzequation} with \(\mathbb{Z}_{2}\) symmetry \(\theta^{1}\leftrightarrow\theta^{2}\) resulting in the degenerate structure of the \(\Gamma^{\alpha}_{\beta\gamma}\) tensors
\begin{equation}
    \Gamma^{1} = \left(
        \begin{array}{cc}
       \Gamma^{1}_{11} & \Gamma^{1}_{12} \\ \Gamma^{1}_{12} & \Gamma^{1}_{22}
        \end{array}\right), \hspace{2mm}\Gamma^{2} = \left(
        \begin{array}{cc}
       \Gamma^{1}_{22} & \Gamma^{1}_{12} \\ \Gamma^{1}_{12} & \Gamma^{1}_{11}
        \end{array}\right)
\label{eq:degeneratenonlinear}
\end{equation}
and \(D=D^{T}=D^{\tau}\), \(\Delta=\Delta^{T}=\Delta^{\tau}\) where \(\tau\) denotes the off-diagonal transpose. It is insightful to transform Eq.~\eqref{equation:mckpzequation} to its normal modes where the rotated fields, denoted by tildes, transform as \(\tilde{\theta}^{1} = {(\theta^{1}+\theta^{2})}/{\sqrt{2}}\), \(\tilde{\theta}^{2} = {(\theta^{1}-\theta^{2})}/{\sqrt{2}}\). These coordinates capture the joint phase and phase difference between the condensates. The linear equations decouple with diffusion matrices \(\tilde{D}_{11} = D_{11}+D_{12}\) and \(\tilde{D}_{22} =  D_{11}-D_{12}\). The noise covariance inherits the diagonal structure \(\langle \tilde{\zeta}^{\alpha} (x,t) \tilde{\zeta}^{\beta} (x',t')\rangle = \delta^{\alpha\beta} (\Delta_{11} \pm \Delta_{12})\delta(x-x')\delta(t-t')\) where \(+,\,-\) are for \((1,1)\) and \((2,2)\) components respectively. Interaction vertices are rank-(1,2) tensors, transforming non-trivially \(\tilde{\Gamma}^{1}_{11} = (\Gamma^{1}_{11} + 2\Gamma^{1}_{12} + \Gamma^{1}_{22})/\sqrt{2}\), \( \tilde{\Gamma}^{1}_{22} = (\Gamma^{1}_{11} - 2\Gamma^{1}_{12} + \Gamma^{1}_{22})/\sqrt{2}\), \(\tilde{\Gamma}^{2}_{12} = (\Gamma^{1}_{11} - \Gamma^{1}_{22})/\sqrt{2}\) while all other non-linear terms vanish. In these coordinates, the \(\mathbb{Z}_{2}\)-theory maps onto 
\begin{equation}
\begin{aligned}
    \partial_{t} \tilde{\theta}^{1} &= \tilde{D}_{11} \partial_{x}^{2} \tilde{\theta}^{1} + \dfrac{1}{2}\left( \tilde{\Gamma}^{1}_{11} (\partial_{x} \tilde{\theta}^{1})^{2} + \tilde{\Gamma}^{1}_{22} (\partial_{x} \tilde{\theta}^{2})^{2} \right) + \tilde{\zeta}_{1} \\ 
    \partial_{t} \tilde{\theta}^{2} &= \tilde{D}_{22} \partial_{x}^{2} \tilde{\theta}^{2} + \tilde{\Gamma}^{2}_{12} (\partial_{x} \tilde{\theta}^{1})(\partial_{x} \tilde{\theta}^{2})  + \tilde{\zeta_{2}}
\end{aligned}
\label{eq:kardarertasequations}
\end{equation}
previously studied by Erta\c{s} and Kardar in dynamic roughening of directed lines \cite{PhysRevLett.69.929}. For decoupled \(\mathbb{Z}_{2}\) symmetric KPZ equations with \(v_{c}=v_{d}=0\), the bare parameters transform to \(\tilde{\Gamma}^{1}_{11}=\tilde{\Gamma}^{1}_{22}=\tilde{\Gamma}^{2}_{12}\) which corresponds to the point \((1,1)\) in FIG.~\ref{fig:combined}(iii). Tuning the intercomponent interactions allows you to move away from the decoupled point into other quadrants, each representing distinct physical behaviours. At large times, the dynamics is dominated by the non-linearity and stability analysis shows that if \(\tilde{\Gamma}^{1}_{22}\) and \(\tilde{\Gamma}^{2}_{12}\) do not have the same sign, there is an instability. This instability arises from the non-hyperbolicity of the linearised equations and non-linear dominated dynamics in Quadrants II and IV in FIG.~\ref{fig:combined}. Tuning \((\tilde{v}_{c},\tilde{v}_{d})\), we can access this instability in Quadrant IV from the SCGLE by choosing bare parameters which satisfy Eq.~\eqref{eq:conditionsstability} but \emph{violate} conditions
\begin{subequations}
\begin{align}
    -1 + \dfrac{\mathcal{R}(u)}{\mathcal{R}(k)}(1+\tilde{v}_{d}) &< \tilde{v}_{c} <  1 - \dfrac{\mathcal{R}(u)}{\mathcal{R}(k)}(1-\tilde{v}_{d}) \label{eq:hyperbolic_1a} \\
    1 - \dfrac{\mathcal{R}(u)}{\mathcal{R}(k)}(1-\tilde{v}_{d}) &< \tilde{v}_{c} <  -1 + \dfrac{\mathcal{R}(u)}{\mathcal{R}(k)}(1+\tilde{v}_{d}) \label{eq:hyperbolic_1b}
\end{align}
\label{eq:conditionnonlinear}
\end{subequations}
for \(\mathcal{R}(u)<\mathcal{R}(k)\) and for \(\mathcal{R}(u)>\mathcal{R}(k)\) respectively. For clarity, we draw attention to the significance of this quadrant in the context of the original \(\mathbb{Z}_{2}\) parameters. In FIG.~\ref{fig:combined}(iii), Quadrant IV is achieved when both X = \(\tilde{\Gamma}^{2}_{12}/\tilde{\Gamma}^{1}_{22} < 0 \) and \( Y = (\tilde{\Gamma}^{1}_{22}\tilde{\Delta}_{22}\tilde{D}_{11})/(\tilde{\Gamma}^{1}_{11}\tilde{\Delta}_{11}\tilde{D}_{22}) > 0 \). In the specific case of the SCGLE mapping, \(\tilde{\Gamma}^{1}_{22}=\tilde{\Gamma}^{1}_{11}\) and \(\tilde{\Delta}_{11}, \tilde{\Delta}_{22}>0\) are always satisfied,  leading to the equivalence of conditions \(Y>0\) and Eq.~\eqref{eq:conditionsstability}, both of which ensure the positive definiteness of the diffusion constants \(\tilde{D}_{11},\,\tilde{D}_{22}>0\). Therefore, Quadrant IV emerges by tuning the effective KPZ interactions, \(\tilde{\Gamma}^{2}_{12} \sim (\Gamma^{1}_{11}-\Gamma^{1}_{22})\) and \(\tilde{\Gamma}^{1}_{11} \sim (\Gamma^{1}_{11}+\Gamma^{1}_{22})\), to have different signs. Consequently, in the \(\mathbb{Z}_{2}\) coordinates, this occurs when the KPZ intercomponent interaction between the phase modes is larger than the KPZ intracomponent interaction \(|\Gamma^{1}_{22}| > |\Gamma^{1}_{11}|\). In this quadrant, the non-hyperbolicity gives rise to large phase differences between nearest-neighbour sites. At the physical level, this triggers the proliferation of STVs, non-trivial windings in the phase due to compactness. The vortex turbulent regime dramatically modifies the scaling, resulting in growth exponent \(\beta=1/2\) and dynamical exponent \(z=1\) as shown in FIG.~\ref{fig:figurestv}. An STV phase has been observed in the single component case in the large noise regime \cite{PhysRevLett.118.085301}, however this phase transition originates from tuning the intercomponent interaction \(\tilde{v}_{c}\). More detail on the analysis of the STV phase can be found in the supplemental material Ref.~\cite{supp}.  

The RG flows are best interpreted in the normal mode basis in adimensionalised coordinates from Eq.~\eqref{eq:kardarertasequations}, closing under the \(\mathbb{R}^{4}\)-parameter space: \(T=\tilde{D}_{22}/\tilde{D}_{11}\), \(X=\tilde{\Gamma}^{2}_{12}/\tilde{\Gamma}^{1}_{11}\), \(Y = (\tilde{\Gamma}^{1}_{22}\tilde{\Delta}_{22}\tilde{D}_{11})/(\tilde{\Gamma}^{1}_{11}\tilde{\Delta}_{11}\tilde{D}_{22})\), \(Z= (\tilde{\Delta}_{11}(\tilde{\Gamma}^{1}_{22})^{2})/(4 \Lambda \tilde{D}_{11}^{3})\) where \(\Lambda\) is the UV cutoff regularising the effective theory. The cutoff has a natural origin in exciton-polariton systems since excitons typically have a Bohr radius up to \(100 \text{\AA}\) giving \(k_{max} = h/100\text{\AA}\). The noise ratio \(W = \tilde{\Delta}_{22}/\tilde{\Delta}_{11}\) is required to be positive definite to retain its connection to probability. The RG flow equations are autonomous ODEs describing how the parameters change as we iteratively integrate out high momentum modes. The boundary condition for the flows is specified at \(\ell=0\) where the parameters assume their initial bare values. The projected flows of \(X\) and \(Y\) are shown in FIG.~\ref{fig:combined}(iii). The parameters \(X,\,Y\) do not change sign under RG, implying that the flows do not cross between quadrants. To understand the phase dynamics, we can concentrate on Quadrants I and IV since the other quadrants present a physical instability from Eq.~\eqref{eq:conditionsstability}. The RG suggests that the scaling should be Edwards-Wilkinson \(z=2\) for the \(\tilde{\theta}_{2}\) mode and KPZ \(z=3/2\) for \(\tilde{\theta}_{1}\) in Quadrant IV, however we must discount it based on the instability from conditions \eqref{eq:hyperbolic_1a} and \eqref{eq:hyperbolic_1b}. Physically, this quadrant as previously discussed corresponds to a disordered STV phase with \(z=1\) for both modes. At this point, we focus on the analysis of the coupled SCGLE in Quadrant I. 

\paragraph{Fluctuation-Dissipation Line} - One important characteristic of the coupled KPZ is that, in contrast to the single component case, there is no longer an incidental Gaussian stationary measure in dimension \(d=1\). A stationary measure arises from a time-independent solution to the Fokker-Planck equation. Ignoring the non-linearity, the Gaussian stationary measure with correlations \(\langle\partial_{x}\theta^{\alpha}(x)\partial_{x'}\theta^{\beta}(x')\rangle = C^{\alpha\beta} \delta(x-x')\) must satisfy the Lyapunov condition \(D C + C D^{T} = 2\Delta\) where \(C=C^{T}\), \(\Delta>0\) and \(D+D^{T}>0\). The condition that the stationary measure remains time-stationary in the presence of interactions requires that the rescaled interaction vertices \(\hat{\Gamma}^{\alpha} = (C^{-1})_{\alpha\beta} \Gamma^{\beta}\) satisfy the cyclicity condition \(\hat{\Gamma}^{\alpha}_{\beta \gamma}=\hat{\Gamma}^{\alpha}_{\gamma\beta}=\hat{\Gamma}^{\beta}_{\gamma\alpha}\) \cite{10.1016/j.jfa.2017.05.002, Roy_2024}. In Quadrants I and III, the flows project onto the fixed line \(X=Y\) where the Fluctuation Dissipation Relation (FDR) is satisfied \(\tilde{\Gamma}^{1}_{22} \tilde{\Delta}_{22}\tilde{D}_{11} = \tilde{\Gamma}^{2}_{12}\tilde{\Delta}_{11}\tilde{D}_{22}\).
On this submanifold, stationarity enforces \(\chi=1/2\) for both components as the measure coincides with the non-interacting theory. Additionally, the one-loop RG predicts KPZ scaling \(z=3/2\) for both components. Furthermore, numerical integration of initial bare parameters in Quadrant I show that the long-time and large system size dynamics converge to the expected exponents \(\beta=1/3\) and \(\chi=1/2\). Along the fixed line, same-time correlations \( S^{\alpha\beta}(|\mathbf{x}-\mathbf{x}'|) = \langle \left(\theta^{\alpha}(\mathbf{x}+\mathbf{x}')-\theta^{\beta}(\mathbf{x}) \right)^{2}\rangle = \tilde{C}^{\alpha\beta} |\mathbf{x}-\mathbf{x}'|^{2\chi}\) are given by two-sided white noise processes with \(\tilde{C}^{\alpha\beta}=[D^{-1}\Delta]^{\alpha\beta}\). Under RG, \(\tilde{C}(l)\) is renormalised until it reaches the FDR line. On the FDR line, emergent time-reversal symmetry enforces the non-renormalisation \(\tilde{C}(l) = \tilde{C}_{0}|_{X=Y}\) \cite{PhysRevE.84.061128}. Consequently, even in the case when \(\tilde{C}_{0}=\mathds{1}\), for \(\tilde{\theta}_{1},\,\tilde{\theta}_{2}\) modes initialised away from \(X=Y\), the respective long-time two-point correlations may not coincide.
\begin{figure}[htbp]
    \centering
    \hspace{-0.75cm}
    \includegraphics[width=0.5\textwidth]{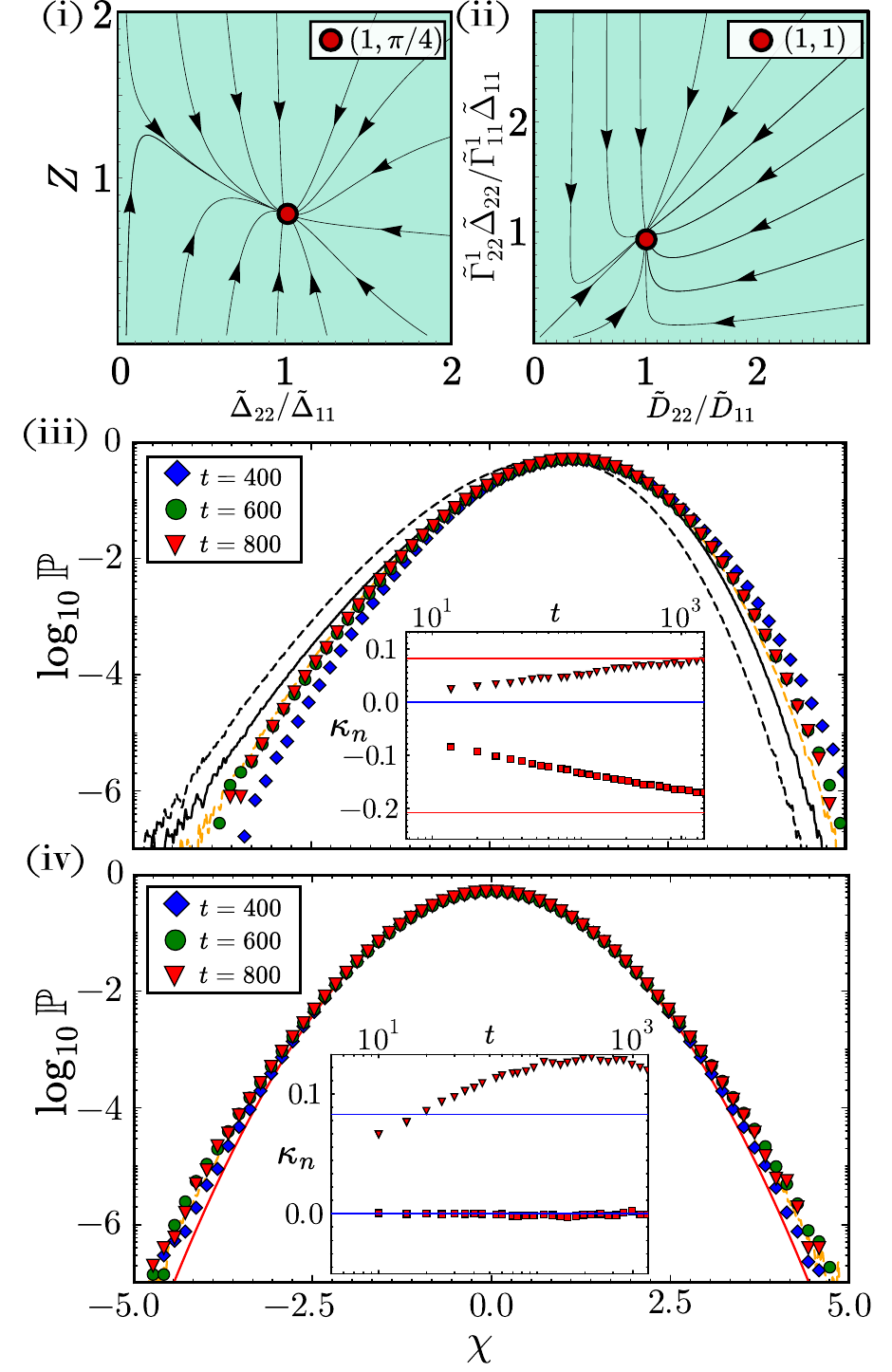}
    \vspace{-0.25cm}
    \caption{Behaviour on the Cole-Hopf line \(X=1\) : (i) RG flows in the \((W,Z)\) plane for initial condition \(T=1,\, Y=W\) flow to decoupled KPZ in the \(\mathbb{Z}_{2}\)-coordinates. (ii) RG flow in the \((T,\tilde{\Gamma}^{1}_{22}\tilde{\Delta}_{22}/\tilde{\Gamma}^{1}_{11}\tilde{\Delta}_{11})\) plane with fixed point \((1,1)\) allowing for a  decoupling transformation. (iii-iv) Distribution of fluctuations \(\chi\) for initial condition \((X,Y) = (1,2)\) (iii) for \(\tilde{\theta}_{1}\) approach the sum of rescaled TW-GOE distributions (yellow). For comparison, we show the normal rescaled TW-GOE distribution \(2^{2/3}F'(-2^{-2/3}\chi_{1}))\) (black dashed) and its translation (black solid) so that the means coincide. (iv) for \(\tilde{\theta}_{2}\) approach the difference of TW-GOE distributions (yellow) as opposed to Gaussian (red). Subfigures within (iii-iv) show time convergence of higher order moments \(\kappa_{n}\) to expected values (horizontal lines) with skewness \(\kappa_{3} = -0.207,\) and kurtosis \(\kappa_{4}=0.0829\) and  \(\kappa_{3} = -0.00,\,\kappa_{4}=0.0829\) respectively for \(\tilde{\theta}_{1}\) and \(\tilde{\theta}_{2}\).}
    \label{fig:figure2}
\end{figure}
\paragraph{Cole-Hopf Line} - A relevant case of the SCGLE for polaritons is where the coherent term \(k_{c}\) dominates over the relaxation term \(k_{d} = 0 \). These parameters map onto the closed \(X=1\) submanifold denoted by the red line in FIG.~\ref{fig:combined}(iii).  A specific sub-case of this is discussed by Funaki et al. \cite{10.1016/j.jfa.2017.05.002} where the diffusion matrix satisfies \(T=1\). The line can be split into two distinct behaviours: (i) \(Y>0\) in Quadrant I where there exists a real Cole-Hopf solution with fixed point coinciding with FDR and (ii) \(Y<0\) in the unstable Quadrant II where there exists a complex Cole-Hopf solution. FIG.~\ref{fig:figure2}(i) shows that a noise coupling between independent KPZ equations is irrelevant. FIG.~\ref{fig:figure2}(ii) details that in the \(X=1\) submanifold in Quadrant I,  \(\tilde{D}_{11}^{*}=\tilde{D}_{22}^{*} \implies T^{*}=1\) at the fixed point and \((\tilde{\Gamma}^{1*}_{22}\tilde{\Delta}_{22}^{*})/(\tilde{\Gamma}^{1*}_{11} \tilde{\Delta}_{11}^{*})=1\). Therefore, the diffusion matrix flows to a form proportional to the identity and the ratio of noises is fixed by the cross-coupling \(\tilde{\Gamma}^{1*}_{22}/\tilde{\Gamma}^{1*}_{11}\). Consequently, at the fixed point, a decoupling transformation exists \(\hat{\theta}^{\alpha}=s^{\alpha}_{\beta}\tilde{\theta}^{\beta}\) with
\begin{equation} s=\begin{pmatrix}\tilde{\Gamma}^{1*}_{11} & (\tilde{\Gamma}^{1*}_{11} \tilde{\Gamma}^{1*}_{22})^{1/2}\\ \tilde{\Gamma}^{1*}_{11}&-(\tilde{\Gamma}^{1*}_{11}\tilde{\Gamma}^{1*}_{22})^{1/2}
\end{pmatrix}
\label{eq:decouplingColeHopf}
\end{equation}
where all parameters assume their renormalised values. This leads to two equivalent, but decoupled, KPZ equations at the fixed point in the coordinate system from transformation Eq.~\eqref{eq:decouplingColeHopf}. For flat initial conditions \(\hat{\theta}(x,0)=0\), the limiting form of the phase is \(\hat{\theta}(x,t)=v_{\infty} t + (M t)^{1/3} \chi \), where \(v_{\infty}\) is the asymptotic growth velocity, \(M\) is a non-universal scaling parameter depending on renormalised parameters, and \(\chi\) is a random variable sampled from a Tracy-Widom (TW) GOE distribution. The fluctuations in the normal mode coordinates have an additional scaling parameter \(\tilde{\Gamma}^{1*}_{22}\) with asymptotic forms 
\begin{equation}
\begin{aligned}
    \tilde{\theta}_{1}(x,t)&=\dfrac{1}{2\tilde{\Gamma}^{1*}_{11}}\left[2v_{\infty}t +(M t)^{1/3}  \left(\chi_{1}+\chi_{2}\right)\right]\\
    \tilde{\theta}_{2}(x,t)&= \dfrac{1}{(\tilde{\Gamma}^{1*}_{11}\tilde{\Gamma}^{1*}_{22})^{1/2}}\left[(M t)^{1/3}\left(\chi_{1}-\chi_{2}\right)\right].
\end{aligned}
\end{equation}
\(\tilde{\theta}_{2}\) has symmetric fluctuations, but unlike a Gaussian, has excess kurtosis shown in the extended tails in FIG.~\ref{fig:figure2}(iv). The kurtosis converges to the expected values at large times. After numerical integration of the multicomponent KPZ with bare parameters \(X=1, \, Y=2\), we rescale the variance of the fluctuations to match a TW-GOE with cumulative distribution \(F(-2^{-2/3} x)\) and \(\langle \chi^{2}\rangle_{c} = 0.638\). The distributions of \(\tilde{\theta}_{1},\,\tilde{\theta}_{2}\) are \(\chi_{1}+\chi_{2}\) and \(\chi_{1}-\chi_{2}\) up to a translation of mean and rescaling of variance. Away from the \(X=1\) line, there is no obvious decoupling transformation, which makes calculating the analytical form of the fluctuation distribution difficult. Nevertheless, \(\tilde{\theta}_{1}\) retains an asymmetric profile and \(\tilde{\theta}_{2}\) retains its inversion symmetry. The distributions can at most depend on two additional parameters \(T, X\). Moving along \(X=Y\) starting from \(0\) at finite time, we see a crossover from Gaussian in \(\tilde{\theta}_{2}\) at the decoupled point \((0,0)\) to more kurtotic distributions. Experimentally and numerically, this showcases that for the multicomponent SCGLE when \(k_{d}=0\), the distribution of fluctuations still has valid subclasses which can be exactly calculated to validate non-equilibrium phenomena.

\paragraph{Conclusion} - We have demonstrated that under RG flow, in the regime of stability of the \(d=1\) SCGLE, we recover a stationary measure, enforcing \(\chi=1/2\) and \(z=3/2\) exponents for both phase components. An interesting case for polaritons arises when \(k_{d}=0\), where the mapped bare parameters lie on the Cole-Hopf line where FDR is restored, and an exact decoupling transformation exists at the fixed point. Our findings reveal the rich physics of multicomponent systems, exhibiting phase-separation in Quadrants II and III due to the onset of gapless mode instabilities, and a novel non-thermal STV regime in Quadrant IV, characterised by dramatically altered dynamical exponents. This phase can be accessed by tuning the intercomponent interactions into the
unstable regime, providing an opportunity to experimentally realise a vortex-dominated phase even in low-noise conditions. This fully determines the phase diagram of the SCGLE and has repercussions for our understanding of multicomponent KPZ systems more broadly. 

\paragraph{Acknowledgements} - The authors acknowledge useful discussions with Herbert Spohn for general insight into coupled KPZ, particularly the hyperbolicity conditions. We gratefully acknowledge financial support from Engineering and Physical Science
Research Council: Grants No. EP/V026496/1 and No. EP/S019669/1 (M.H.S. and P.C.), EP/S021582/1 (P.C), and EP/T517793/1 (H.W.). The authors acknowledge the use of the UCL HPC facilities and associated support services in the completion of this work.

\bibliographystyle{apsrev4-1} 
\bibliography{bib} 

\end{document}


\UseRawInputEncoding

\title{Multicomponent Kardar-Parisi-Zhang Universality in Degenerate Coupled Condensates Supplemental Materials}

\author{H.~Weinberger\textsuperscript{1}}
\author{P.~Comaron\textsuperscript{1,2}}
\author{M.H.~Szyma\'nska\textsuperscript{1}}
\affiliation{\textsuperscript{1}Department of Physics and Astronomy, University College London, 
Gower Street, London, WC1E 6BT, United Kingdom}
\affiliation{\textsuperscript{2}CNR NANOTEC, Institute of Nanotechnology, Via Monteroni,  73100 Lecce, Italy}

\begin{abstract}
    In this supplemental material, we outline the Wilsonian renormalisation group procedure for the multicomponent degenerate KPZ equations and other numerical methods used to produce the results presented in the main letter. The material is divided into five subsections: (A) The derivation of the multicomponent \(\mathbb{Z}_{2}\) KPZ equation from Keldysh field theory and discussion around its stability (B) Properties of the multicomponent KPZ equation, particularly its adimensionalised form and conditions for a stationary measure (C) Derivation and discussion of the Wilsonian RG procedure from the MSRJD functional integral, along with subsequent discussion of the flow equations (D) Direct numerical integration of the KPZ equation and comparison with Tracy-Widom GOE distributions including additional simulations of the \((X,Y)\) plane in FIG.~\ref{fig:combined}(iii) (E) Direct numerical integration of the SCGLE, focusing on compact features such as spacetime vortices proliferating in the instability regime in Quadrant IV. 
\end{abstract}

\maketitle

\onecolumngrid
\vspace{-1cm}
\section{ A. Weakly Interacting Driven-Dissipative Condensates}
In this section, we derive the effective theory for the Nambu-Goldstone (NG) modes of the \(\mathbb{Z}_{2}\) coupled condensate under drive and dissipation using the Lindbladian formalism and Keldysh field theory. Starting from the Lindbladian, we motivate the SCGLE as a semiclassical description of a quantum model suitable for weakly-interacting Bosons at mesoscopic scales. 
\subsection{Schwinger-Keldysh Partition Function and Semiclassical Action}
In equilibrium zero-temperature QFTs, we are typically concerned with calculating time ordered expectations of operators in the ground state \(\ket{\Omega}\). Non-equilibrium QFTs, such as Keldysh field theory, share similarities but now instead of working with states \(\ket{i}\), the theory focuses on the dynamical evolution of the density matrix \(\hat{\rho}(-\infty)\). The fundamental object is the partition function \(\mathcal{Z}= \text{Tr}[e^{\mathcal{\hat{L}}t}[\hat{\rho}(-\infty)]] = 1\) which can be used to calculate n-point correlation functions. The partition function equals unity in the absence of sources, as the Lindbladian operator is a completely positive trace-preserving map. The Lindbladian \(\hat{\mathcal{L}}\in \mathrm{End}(\mathcal{H}^{*}\otimes\mathcal{H})\) serves as the generator of time evolution with dynamical semigroup structure for the space of density matrices with Markovian coupling to a bath reservoir. In our case, the density matrix \(\hat{\rho}\) evolves under the coherent Hamiltonian 
\begin{equation}
\begin{aligned}
    \hat{H} =  \int_{x} \Bigg[&\sum_{i}\left[k_{c}(\grad{\hat{\Psi}}_{i})^{\dag} (\grad{\hat{\Psi}}_{i})) + \mu_{c} ( \hat{\Psi}_{i}^{\dag}\hat{\Psi}_{i})\right] +\sum_{i=j} \left[u_{c}\hat{\Psi}^{\dag}_{i}\hat{\Psi}^{\dag}_{j}\hat{\Psi}_{j}\hat{\Psi}_{i}\right]  + \sum_{i\neq j}  \left[v_{c}\hat{\Psi}^{\dag}_{i}\hat{\Psi}^{\dag}_{j}\hat{\Psi}_{j}\hat{\Psi}_{i}\right]
\end{aligned}
\end{equation}
where the operators are normal ordered. Assuming that the interactions have no spatial extent, there are local intracomponent \(u_{c}\) and intercomponent \(v_{c}\) \(4\)-point density-density interactions. The operators obey the same-time algebraic structure \([\hat{\Psi}_{i}(x,t),\hat{\Psi}^{\dag}_{j}(x',t)]=\delta_{ij}\delta(x-x')\) and \([\hat{\Psi}_{i}(x,t),\hat{\Psi}_{j}(x',t)]=0\). A \(\mathbb{Z}_{2}\) symmetry is imposed, requiring invariance under inversions in field space \(\hat{\Psi}_{1} \leftrightarrow \hat{\Psi}_{2}\). To derive the mesoscopic Eq.~(\ref{eq:coupled_SCGLE}), we select the following Lindblad operators with homogeneous constants (i) single particle pumping from lasing \(\hat{\Psi}^{\dag}_{i}\) proportional to \(\gamma_{p}\) (ii) single particle losses to the excitonic reservoir \(\hat{\Psi}_{i}\) proportional to \(\gamma_{l}\) (iii) two particle non-linear losses \(\hat{\Psi}_{i}\hat{\Psi}_{i}\) proportional to \(2 u _{d}\) (iv) cross-component two particle non-linear losses \((\hat{\Psi}_{1}\hat{\Psi}_{2} + \hat{\Psi}_{2}\hat{\Psi}_{1}) \) proportional to \(2 v_{d}\). This preserves the \(U(1)\times U(1)\) symmetry of the action, ensuring that the low-energy sector is described by two coupled gapless NG modes. From the dynamical semigroup structure of the Lindbladian and the Schwinger-Keldysh closed loop formalism, the real-time field theory is constructed by resolving coherent states in Trotter steps
\begin{equation}
    \int \prod_{i}\left(\dfrac{d\psi_{i,t_{n}} d\bar{\psi}_{i,t_{n}}}{\pi}\right) e^{-\sum_{i}|\psi_{i,t_{n}}|^{2}} \ket{\Psi_{t_{n}}}\bra{\Psi_{t_{n}}} = \mathds{1}, \;\;\; \bra{\Psi_{t_{l}}}\ket{\Psi_{t_{m}}} = e^{\sum_{i}\bar{\psi}_{i,t_{l}}\psi_{i,t_{m}}}
\end{equation}
where the operators are indexed by \(t_{n}\) on the \emph{forward} and \emph{backward} contour. The field operators act on coherent states as \(\hat{\Psi}_{i} \ket{\Psi_{t_{n}}} = \psi_{i,t_{n}}\ket{\Psi_{t_{n}}}\) with \(\psi_{i,t_{n}}\in \mathbb{C}\). Taking Trotter time \(\Delta t\) to zero gives the coherent state path integral. There is a doubling of the field degrees of freedom and truly non-equilibrium dynamics arises from couplings between fields on the \emph{forward} contour \(C_{+}\) running from \(t=-\infty\) to \(t=\infty\) and the \emph{backward} contour \(C_{-}\) running from \(t=\infty\) to \(t=-\infty\). For further details, see the review article by Sieberer et al. \cite{Sieberer_2016}. In the semiclassical limit---robustly justified in the vicinity of a phase transition---terms greater than \(\mathcal{O}((\psi^{q})^{2})\) are ignored, giving the partition function
\begin{equation}
    \mathcal{Z} = \int \mathcal{D}[\psi^{c}_{i},\psi^{q}_{i}] \hspace{1mm} e^{i\mathcal{S}[\psi_{i}^{c},\psi_{i}^{q}]}
\label{seq:partitionfunction}
\end{equation}
with microscopic action 
\begin{equation}
\begin{aligned}
    \mathcal{S} &= \int_{x,t} \left(\left[\bar{\psi}^{q}_{i}\left(i\partial_{t} + k\grad^{2} - \mu \right)\psi^{c}_{i}\right] -  u\left[\bar{\psi}_{i}^{q}\bar{\psi}_{i}^{c}\psi_{i}^{c}\psi_{i}^{c}\right] - v \left[\bar{\psi}_{1}^{c}\psi_{1}^{c}\bar{\psi}_{2}^{q}\psi_{2}^{c} + (1 \leftrightarrow 2)\right] + \text{c.c.}\right) \\
    & + 4i u_{d} \bar{\psi}^{c}_{i}\psi^{c}_{i}\bar{\psi}^{q}_{i} \psi^{q}_{i} + 2 i v_{d} \left[\bar{\psi}^{c}_{2}\psi^{c}_{2}\bar{\psi}^{q}_{1} \psi^{q}_{1} + \bar{\psi}^{c}_{1}\psi^{c}_{2}\bar{\psi}^{q}_{2} \psi^{q}_{1} + (1 \leftrightarrow 2)\right] + 2 i \gamma_{t} \bar{\psi}^{q}_{i}\psi^{q}_{i}
\end{aligned}
\label{eq:actionfullaction}
\end{equation}
where \(k = k_{c} - i k_{d}\), \(\mu = \mu_{c} + i \mu_{d}\), \(u=u_{c} - i u_{d}\), \(v=v_{c}-iv_{d}\) with free indices implicitly summed over. In the original Lindblad parameters, \(\mu_{d} = (\gamma_{p}-\gamma_{l})/2\) and \(\gamma_{T} = (\gamma_{p}+\gamma_{l})/2\). When the lasing gain \(\gamma_{p}\) exceeds the single particle losses to the excitonic bath modes \(\gamma_{l}\), spontaneous symmetry breaking (SSB) of the \(U(1)\times U(1)\) symmetry occurs. Since the theory is Gaussian in the quantum fields, auxiliary white noise fields \(\zeta_{i} \in \mathbb{C}\) can be introduced via the Hubbard-Stratonovich transformation. After integrating out quantum fields and absorbing constants into the functional measure, we obtain the SCGLE. However, there is ambiguity in the form of the noise covariance for Eq.~(\ref{eq:coupled_SCGLE}) as including density contributions gives
\begin{equation}
    \langle \zeta_{i}(x,t) \zeta_{j}(x',t')\rangle = \begin{pmatrix}2((2 |\psi^{c}_{1}|^{2}u_{d} +  |\psi^{c}_{2}|^{2}v_{d}) + \gamma_{T} ) && 2 \bar{\psi}^{c}_{2}\psi^{c}_{1}v_{d}\\2 \bar{\psi}_{1}^{c}\psi_{2}^{c} v_{d}&&2((2 |\psi^{c}_{2}|^{2}u_{d} +  |\psi^{c}_{1}|^{2}v_{d}) + \gamma_{T} ) \end{pmatrix}_{ij} \; \delta(x-x')\, \delta(t-t').
\label{seq:covariance_full}
\end{equation}
For the direct integration of the SCGLE in FIG.~\ref{fig:combined}(ii) and analysis of the onset of spacetime vortices detailed in Section E, we assume that the noise is additive consistent with the density correction being subleading. In this approximation, the covariance becomes diagonal and equal to \(2\gamma_{T}\) giving Eq.~(\ref{eq:coupled_SCGLE}).  
\subsection{Diffusive NG Modes Stability}
Starting from the SCGLE and ignoring the additive white noise, we can derive Eq.~\eqref{eq:conditionsstability} for stability of the Bogoliubov fluctuations around a homogeneous mean-field. This procedure reveals two gapped density modes and two NG diffusive modes, unlike the equilibrium case. In the SSB phase, fluctuations around the mean-field \(\rho=|\psi^{(0)}_{i}|^{2}\) can be considered. Guided by the coset construction, the density-phase representation parameterisation
\begin{equation}
    \psi_{i} = \sqrt{\rho + \chi_{i}(x,t)}e^{i(\theta_{i,0}+\theta_{i}(x,t))}
\end{equation}
will correctly capture the low energy theory where \(\theta_{i,0},\, \theta_{i}, \, \rho, \, \chi_{i}\in \mathbb{R}\). Inserting the ansatz into Eq.~ (\ref{eq:coupled_SCGLE}) and considering terms up to linear order in fluctuations gives the linear equation
\begin{equation}
    \partial_{t}\left(\begin{array}{c}\chi_{1}\\ \theta_{1} \\ \chi_{2} \\ \theta_{2}\end{array}\right)=\left(
\begin{array}{cccc}
  k_{d} \partial_{x}^2 - 2 \rho u_d  & -2 \rho k_{c} \partial_{x}^2 & -2
   \rho v_d & 0 \\
 - k_{c}\partial_{x}^2 + 2 \rho  u_c & - 2\rho  k_{d} \partial_{x}^2& 2\rho v_c & 
   0 \\
 -2 \rho v_d & 0 &  k_{d} \partial_{x}^2 - 2 \rho u_d  & -2 \rho k_{c} \partial_{x}^2 \\
 2 \rho v_c  & 0 & - k_{c}\partial_{x}^2 + 2 \rho u_c & - 2 \rho  k_{d} \partial_{x}^2\\
\end{array}
\right)\left(\begin{array}{c}\chi_{1}\\ \theta_{1} \\ \chi_{2} \\ \theta_{2}\end{array}\right).
\end{equation}
We define the Fourier transform convention 
\begin{equation}
    \theta^{\alpha}(x,t) = \int_{k,\omega}\theta^{\alpha}_{\omega, k}  e^{-i \omega t + i k\cdot x}
\label{seq:fouriertransformconvention}
\end{equation} 
with integration measure \(d\omega/(2\pi)\). Transforming the fields to Fourier space gives a linear equation \(-i\omega \Psi = M \Psi \). For stability, fluctuations must decay, requiring that all eigenvalues lie in the lower-half complex plane. Expanding to \(O(k^{2})\), the eigenvalues are 
\begin{equation}
\begin{aligned}
    \omega_{\chi,\pm}&= -i \left[2\rho(u_{d}\pm v_{d}) -  k^2 \left( k_{c} \dfrac{u_{c}\pm v_{c}}{u_{d}\pm v_{d}}-k_{d}\right)\right] \\ \omega_{\theta,\pm}&= - i k^2 \left( k_{c} \dfrac{u_{c}\pm v_{c}}{u_{d}\pm v_{d}}+k_{d}\right).
\label{seq:frequencies_bogoliubov}
\end{aligned}
\end{equation}
As \(k\rightarrow 0\), \(\omega_{\chi,\pm}\) has finite damping thus is gapped. For zeroth order stability, \(u_{d}-v_{d}>0 \implies \tilde{v}_{d}= v_{d}/u_{d}<1\); the intercomponent non-linear dissipative term is bounded by the intracomponent non-linear dissipative term. If this condition is violated, the gapped modes become unstable and drive a dynamical phase transition. Stability of the gapless modes while in the regime \(u_{d}-v_{d}>0\) imposes that both 
\begin{equation}
    \left( k_{c} (u_{c}+v_{c})+k_{d}(u_{d}+v_{d})\right) > 0 \; \mathrm{and}  \left(  k_{c} (u_{c}+v_{c})+k_{d}(u_{d}+v_{d})\right) > 0 
\end{equation}
are satisfied. Assuming that we are the positive mass regime \(k_{c}>0\), the intracomponent interaction is repulsive \(u_{c}>0\) gives Eq.~\eqref{eq:conditionsstability} 
\begin{equation}
    -1 - \mathcal{R}(u)\mathcal{R}(k) (1+\tilde{v}_{d})\tilde{v}_{c} < 1 + \mathcal{R}(u)\mathcal{R}(k) (1-\tilde{v_{c}})
\end{equation}
where \(\mathcal{R}(k)=k_{d}/k_{c}\) and \(\tilde{v}_{c}=v_{c}/u_{c}\). Approaching the equilibrium limit \(R(k)\rightarrow 0\), \(R(u)\rightarrow 0\), \(\tilde{v}_{d}\rightarrow 0\) recovers the more restrictive equilibrium condition \(-1<\tilde{v}_{c}<1\) and the four modes become gapless. This means that non-equilibrium condensates are more stable against gapless diffusive fluctuations. In Quadrants II and III in FIG.~\ref{fig:combined}(iii), Eq.~\eqref{eq:conditionsstability} is not satisfied and this drives a dynamical phase transition and subsequent fragmentation of the condensate. An equivalent formulation of Eq.~\eqref{eq:conditionsstability} can be obtained by demanding that the diffusion matrix for the mapped KPZ effective equation is positive-definite.
\subsection{Low Energy Effective Field Theory}
The low-energy sector dynamics, described by the multicomponent KPZ Eq.~\((\ref{equation:mckpzequation})\), can be found using the density-phase representation, but in the enlarged Keldysh space. Due to couplings of fields lying on forward and backward branches in Action (\ref{eq:actionfullaction}), the symmetries along the forward and backward contours are explicitly broken down to the diagonal subgroup \(U(1)\times U(1)\) with action on the complex fields 
\begin{equation}
    \psi^{c}_{i} \mapsto e^{i\theta_{i}} \psi^{c}_{i} , \; \; \; \psi^{q}_{i}\mapsto e^{i\theta_{i}} \psi^{q}_{i} 
\end{equation}
where \(i\) subscripts index the component. In the condensed phase \(\rho>0\), the NG modes are parameterised \(\psi^{c}_{i} = \sqrt{\rho_{i}} e^{i\theta_{i}}\) and \(\psi^{q}_{i}=\xi^{R}_{i} + i \xi^{I}_{i}\). The Jacobian of this coordinate transformation is absorbed into the path integral measure since it is field independent and poses no issues for \(\rho_{i}>0\). The new fields have target spaces: the entire \(\mathbb{R}\) line for \(\xi^{R/I}_{i}\),  \(\mathbb{R}^{+}\setminus0\) for \(\rho_{i}\), and the space \([0,2\pi]\) for \(\theta_{i}\). Integrating out gapped density fluctuations around a strong mean-field leads to the effective field theory describing the low-energy sector. The mean-field solution is valid for weakly-interacting Bosons in the condensed phase where fluctuations are small. The low-energy partition function is given by
\begin{equation}
    \mathcal{Z} = \int \mathcal{D}[\rho_{i},\xi^{R}_{i},\xi^{I}_{i},\theta_{i}] \hspace{1mm} e^{i \left(S^{R}[\rho_{i},\xi^{R}_{i},\xi^{I}_{i},\theta_{i}] + i S^{I}[\rho_{i},\xi^{R}_{i},\xi^{I}_{i},\theta_{i}]\right)}
\end{equation}
where terms quadratic in quantum fields give rise to the imaginary part of the action \(S^{I}\)
\begin{equation}
    S^{I} = \int_{x,t} 2(2\rho_{1} u_{d} + \rho_{2} v_{d} + \gamma_{T} )\left[ (\xi^{R}_{1})^{2} +(\xi^{I}_{1})^{2}\right] + (8 \sqrt{\rho_{1}} \sqrt{\rho_{2}} v_{d})\left[\xi^{R}_{1}\xi^{R}_{2}+ \xi^{I}_{1}\xi^{I}_{2}\right] + 1\leftrightarrow2.
\label{seq:noiseterms}
\end{equation}
The integrand is positive-definite ensuring convergence for a contour running along the real axis. Classically motivated, the NG modes should be fluctuations on top of a mean-field homogeneous solution. Finding the time-stationary and spatially homogeneous saddle-points in gapped quantities by taking variations in \(\rho_{i}(x,t)\), we obtain \(\delta_{i} S = 0 \implies \xi^{R/I}_{i} = 0\) is a trivial solution with vanishing quantum fields. Taking variations in quantum fields \(\xi_{i}^{R/I}\) gives the causal saddle-points for the dynamical equations of the retarded operator in the Keldysh action. Although non-trivial saddle-point solutions may exist, they are not considered here. In this approximation, the saddle-point equations become 
\begin{equation}
\begin{gathered}
    \mu_{c} + \rho_{1} u_{c} + \rho_{2} v_{c} = 0,  \;
    \mu_{c} + \rho_{2} u_{c} + \rho_{1} v_{c} = 0 \\
    \mu_{d} - \rho_{1} u_{d} - \rho_{2} v_{d} = 0,  \;
    \mu_{d} - \rho_{2} u_{d} - \rho_{1} v_{d} = 0 \\
\end{gathered}
\end{equation}
respecting the \(\mathbb{Z}_{2}\) symmetry. This fixes the mean-field density and provides a consistency relation determining the chemical potential \(\mu_{c}\). Expanding in fluctuations \(\chi_{i} = \rho_{i}(x,t)-\rho, \, \xi^{R}_{i}, \, \xi^{I}_{i}\) around the mean-field \(\phi_{i,0}=(\rho=\mu_{d}/(u_{d}+v_{d}),0,0)\), the action is expanded up to quadratic order in gapped fluctuations \(\phi_{i} = (\rho_{i},\xi^{R}_{i},\xi^{I}_{i})\) ignoring derivatives of gapped quantities 
\begin{equation}
    S[\theta_{i},\xi^{R}_{i},\xi^{I}_{i},\rho_{i}] = S_{0}[\phi_{i,0},\theta_{i}] + \int_{x} \left.\dfrac{\delta S}{\delta \phi_{i}}\right\vert_{0}(\phi_{i} - \phi_{i,0}) + \dfrac{1}{2}\int_{x} \left.\dfrac{\delta^{2} S}{\delta \phi_{i}\delta\phi_{j}}\right\vert_{0} (\phi_{i}-\phi_{i,0})(\phi_{j}-\phi_{j,0})
\end{equation}
where the Hessian will be an operator on the field fluctuations. Key properties of the expanded action include: (i) \(S_{0}[\phi_{i,0}]|_{0} = 0\) due to vanishing quantum fields at the saddle-point, and (ii) static terms in \(\delta S/\delta \phi_{i}|_{0}\) vanish, while derivatives in \(\theta_{i}\) fields still contribute. This indicates that this is not a true saddle-point but more a classically motivated mean-field solution (iii) the Hessian introduces quadratic couplings in fluctuation fields, easily integrated out owing to their Gaussian structure. The extension to allow \(\chi\) integration limits between \([-\infty,\infty]\) is justified by noting that deviations from the mean-field saddle-point contribute little to the integral. This enables computation of the integral over \(\chi_{i}\) 
\begin{equation}
\begin{gathered}
    \prod_{i}\int D[\chi_{i}] e^{i\int_{x} 2\sqrt{\rho}\chi_{i}\left( v_{c} (\xi^{R}_{i} - \xi^{R}_{j}) + u_{d} \xi_{i}^{I} + v_{d}\xi^{I}_{j}-\xi^{R}_{i}(u_{c}+v_{c})\right)} 
    = \prod_{i} \delta\left( u_{d} \xi^{I}_{i} + v_{d} \xi^{I}_{j} - v_{c} \xi^{R}_{j}- u_{c}\xi^{R}_{i}\right)
\end{gathered}
\label{eq:delta_constraints}
\end{equation}
for \(i\neq j\) and using the integral representation of the delta functional. The integral projects onto a manifold of trajectories where \(\xi^{R/I}_{i}\) fluctuations are constrained by the delta functional conditions (\ref{eq:delta_constraints}). The product of delta functions is non-trivial and demands discussion for carrying out the subsequent integral over \(\xi_{i}^{I}\). We can consider a single point \((x_{0},t_{0})\) noting that the argument of the delta function is time and spatially local and then generalise the argument for the entire integral. This involves integrating an expression like
\begin{equation}
    \int_{U \subseteq \mathbb{R}^{4}} d\xi^{I}_{1}d\xi^{I}_{2} \hspace{1mm} \delta(g(\xi^{R}_{1},\xi^{R}_{2},\xi^{I}_{1},\xi^{I}_{2}))\hspace{1mm}\delta(h(\xi^{R}_{1},\xi^{R}_{2},\xi^{I}_{1},\xi^{I}_{2})) f(\xi^{R}_{1},\xi^{R}_{2},\xi^{I}_{1},\xi^{I}_{2};\theta_{1},\theta_{2})
\end{equation}
where \(g\) and \(h\) are the arguments of the delta functions from Eq.~(\ref{eq:delta_constraints}) and \(f\) is the remaining functional of gapped quantities and phase fields. After the constraints from the delta function, we have an integral over a surface given by \(U \cap (g^{-1}(0) \cap h^{-1}(0))\). This is well-defined if \(g\) and \(h\) are linearly independent. Considering a diffeomorphism \(\phi:V \rightarrow U\) such that \(\mathbf{\xi} \mapsto \phi(\mathbf{\xi}) \in U\) allows for a change of variables
\begin{equation}
    \int_{U \subseteq \mathbb{R}^{4}} d\mathbf{\xi} \hspace{1mm} \delta(g(\mathbf{\xi}))\hspace{1mm}\delta(h(\mathbf{\xi})) f(\mathbf{\xi};\theta_{1},\theta_{2}) \mapsto \int_{V \subseteq \mathbb{R}^{4}} d \mathbf{\xi} \delta(g(\phi(x))) \delta(h(\phi(x))) f(\phi(x);\theta_{1},\theta_{2}) J(\phi(\mathbf{\xi}))
\end{equation}
to a coordinate system where the delta functions act on orthogonal spaces
\begin{equation}
\begin{gathered}
    \tilde{\xi}^{I}_{1} = u_{d}\xi^{I}_{1} +  v_{d}\xi^{I}_{2}, \;\;\tilde{\xi}^{I}_{2} = v_{d}\xi^{I}_{1} +  u_{d}\xi^{I}_{2}. 
\end{gathered}
\end{equation}
This is a valid diffeomorphism provided \(u_{d}\neq v_{d}\) where a density instability arises. The delta functions now act on orthogonal spaces, allowing the integrals to be computed. In the original action in Eq.~(\ref{eq:actionfullaction}), intercomponent density-density interactions \(v_{c},\, v_{d}\) mediate couplings of fields \(\xi^{R/I}_{i}\chi_{\overline{i}}\) between the quantum and classical field fluctuations, while phase fluctuations \(\theta_{i}\) only couple to quantum fields \(\xi_{i}^{R/I}\) within the same component. Integrating out density fluctuation fields \(\chi_{i}\) and \(\xi^{I}_{i}\), gives rise to an effective cross-channel interaction between derivatives of the phase fluctuations and the real quantum fields. The low-energy effective theory has partition function 
\begin{equation}
    \mathcal{Z}= \int D[\boldsymbol{\theta},\tilde{\boldsymbol{\xi}}^{R}] e^{i \int \left[\tilde{\xi}^{R}_{\alpha} \left(\partial_{t}\theta^{\alpha} - D^{\alpha}_{\beta}\grad^{2} \theta^{\alpha} - \Gamma^{\alpha}_{\beta\gamma}(\grad\theta^{\beta})(\grad \theta^{\gamma})\right) + i\ \tilde{\xi}_{\alpha}\Delta^{\alpha}_{\beta} \tilde{\xi}^{\beta} \right]}
\label{eq:kpz_path_integral}
\end{equation}
with rescaled real quantum fields \(\tilde{\xi}^{R}_{\alpha} = -2\sqrt{\rho}\hspace{1mm}\xi_{\alpha}\). At this point, it is customary to introduce an auxiliary field using a Hubbard-Stratonovich transformation leading to the stochastic PDE Eq.~(\ref{equation:mckpzequation}) in It\^{o} discretisation. The KPZ parameters in Eq.~(\ref{eq:kpz_path_integral}) are
\begin{equation}
\begin{gathered}
    D_{11} = k_{c}\mathcal{C}_{1}(u_{d}u_{c} - v_{c}v_{d}) + k_{d} , \, 
    D_{12} = k_{c}\mathcal{C}_{1}(v_{c}u_{d} - u_{c}v_{d}),\\ 
    \Gamma^{1}_{11} = -2 k_{c} + 2k_{d}\mathcal{C}_{1}\left(u_{c}u_{d}-v_{c}v_{d}\right), \, 
    \Gamma^{1}_{22} = 2k_{d}\mathcal{C}_{1}(v_{c}u_{d} - u_{c}v_{d}), \;    \Gamma^{1}_{12}= 0 
\end{gathered}
\end{equation}
where \(\mathcal{C}_{1} = (u_{d}^{2} - v_{d}^{2})^{-1}\). The other coordinates can be inferred from the \(\mathbb{Z}_{2}\) symmetry \(1\leftrightarrow2\). All unspecified elements of the \(\Gamma^{\alpha}_{\beta\gamma}\) tensor are zero. The covariance matrix has a complicated structure, however becomes diagonal in the normal mode basis
\begin{equation}
\begin{gathered}
    \tilde{\Delta}_{11} = \frac{\left(\left(u_c + v_c\right){}^2+\left(u_d+v_d\right){}^2\right) \left(2 \rho  (u_d + v_{d})+\gamma _T\right)}{2 \rho  \left(u_d-v_d\right)^2}, \;
    \tilde{\Delta}_{22} = \frac{\left(\left(u_c - v_c\right){}^2+\left(u_d-v_d\right){}^2\right) \left(2 \rho  u_d+\gamma _T\right)}{2 \rho  \left(u_d-v_d\right)^2}.
\end{gathered}
\end{equation}
If we use the simpler form of the Action in Eq.~(\ref{seq:noiseterms}) ignoring the additional cross-couplings as done in Eq.~(\ref{seq:covariance_full}), the covariance matrix becomes
\begin{equation}
\begin{gathered}
    \tilde{\Delta}_{11} = \gamma_{T}\frac{\left(\left(u_c + v_c\right){}^2+\left(u_d+v_d\right){}^2\right) }{2 \rho  \left(u_d-v_d\right)^2}, \;
    \tilde{\Delta}_{22} = \gamma_{T}\frac{\left(\left(u_c - v_c\right){}^2+\left(u_d-v_d\right){}^2\right)}{2 \rho  \left(u_d-v_d\right)^2} 
\end{gathered}
\end{equation}
and \(\rho=\mu_{d}/(u_{d}+v_{d})\) is the mean-field density. In the limit \(v_{c},v_{d} \rightarrow0\) the SCGLE equations become decoupled as expected and agrees with the form presented in Ref.~\cite{Sieberer_2016}. The effective equations retain the \(\mathbb{Z}_{2}\) degeneracy, motivating the symmetry constrained Wilsonian RG.
\section{B. Properties of the Multicomponent KPZ Equation}
Before running the Wilsonian RG, we review concepts around the multicomponent KPZ equation to better interpret the flows. This section discusses the expected scaling form of the multicomponent KPZ solutions, the form of the adimensionalised equations useful for interpreting the RG flows, and the emergence of a stationary probability measure for the KPZ fields in specific submanifolds that satisfy the cyclicity condition for their interaction vertices in Eq.~(\ref{equation:mckpzequation}).
\subsection{Scaling Form and Adimensionalisation}
In the single-component case, KPZ solutions follow a scaling form 
\begin{equation}
    \theta(x,t) \sim  t^{\chi/z}f(x/t^{1/z})
\label{eq:scaling_form}
\end{equation}
where \(\chi=1/2\) is the roughness exponent, indicating that the theory is locally Brownian, and \(z=3/2\) defines the KPZ dynamical exponent. Praehoffer and Spohn exactly determined the stationary KPZ distribution for the two-point correlation function~\cite{Prähofer2002}. This scaling form implies that the transverse scale evolves as \(t^{1/3}\) and the lateral scale evolves as \(t^{2/3}\). For the single-component KPZ equation, we are at liberty to rescale the theory to have one characteristic parameter, 
\begin{equation}
    \partial_{t}\theta = \partial_{x}^{2}\theta + \sqrt{g} (\partial_{x}\theta)^{2} + \sqrt{2} \zeta, \; \sqrt{g} = \dfrac{\lambda \epsilon^{1/2} \Delta^{1/2}}{2D^{3/2}},
\label{eq:KPZ1D}
\end{equation}
controlling the scaling behaviour: \(g^{*}=0\) is the Edwards-Wilkinson (EW) limit where \(z=2\),and \(g^{*}=\pi/2\) is the KPZ fixed point in \(d=1\) predicted by Wilsonian RG. In Eq.~\eqref{eq:KPZ1D}, we have used a different protocol for the white noise with unit covariance, but switching to the covariance matrix definition used in the Eq.~(\ref{equation:mckpzequation}) is done by decomposing the coupled noise into two independent white noise processes. From the rotated KPZ Eq.~(\ref{eq:kardarertasequations}), if \(\tilde{D}^{\alpha\beta}\) is not proportional to \( \mathds{1}_{2\times2}\), we are unable to simply rescale our fields to have unit \(\tilde{C}^{\alpha\beta} = [\tilde{D}^{-1}\tilde{\Delta}]^{\alpha \beta}\) for the rotated two-point correlation matrix of the non-interacting theory. When \(\tilde{D}_{11}\neq\tilde{D}_{22}\), an additional scaling parameter emerges, requiring further investigation. To illustrate this, we derive the adimensionalised equations showcasing four relevant parameters. We construct a family of rescaled solutions indexed by \(\epsilon\) for each field 
\begin{equation}
    \tilde{\theta}_{1}(x,t) = \epsilon^{\chi_{1}} \tilde{\theta}_{1}^{\epsilon}(\epsilon^{-1}x,\epsilon^{-z_{1}}t), \;\; \tilde{\theta}_{2}(x,t) = \epsilon^{\chi_{2}} \tilde{\theta}_{2}^{\epsilon}(\epsilon^{-1}x,\epsilon^{-z_{2}}t)
\end{equation}
where we initially entertain the idea of different scaling parameters \(z_{1},\,z_{2}\) in time, a condition which we will later abandon. In this scenario, the new derivatives are defined with respect to \(x'=\epsilon^{-1}x\) and \(t_{1}=\epsilon^{-z_{1}}t,\,t_{2}=\epsilon^{-z_{2}}t\). The dynamical equations in the rescaled fields become 
\begin{subequations}
\begin{align}
\partial_{t_{1}} \tilde{\theta}^{\epsilon}_{1} &= \epsilon^{z_{1} -2} \tilde{D}_{11} \partial_{x'}^{2} \tilde{\theta}^{\epsilon}_{1} +\dfrac{1}{2}\epsilon^{\chi_{1} + z_{1} -2} \tilde{\Gamma}^{1}_{11} (\partial_{x'}\tilde{\theta}^{\epsilon}_{1})^{2} +\dfrac{1}{2}\epsilon^{2\chi_{2}-\chi_{1}+z_{1}}\tilde{\Gamma}^{1}_{22} (\partial_{x'}\tilde{\theta}^{\epsilon}_{2})^{2} + \epsilon^{z_{1}-\chi_{1}}\dfrac{1}{\sqrt{\epsilon^{z_{1}+1}}}\sqrt{2\tilde{\Delta}_{11}} \zeta^{\epsilon}, \\ 
\partial_{t_{2}} \tilde{\theta}^{\epsilon}_{2} &= \epsilon^{z_{2} -2} \tilde{D}_{22} \partial_{x'}^{2} \tilde{\theta}^{\epsilon}_{2} + \epsilon^{\chi_{1} + z_{2} - 2} \tilde{\Gamma}^{2}_{12} (\partial_{x'}\tilde{\theta}^{\epsilon}_{1}) (\partial_{x'}\tilde{\theta}^{\epsilon}_{2}) + \epsilon^{z_{2}-\chi_{2}}\dfrac{1}{\sqrt{\epsilon^{z_{2}+1}}} \sqrt{2\tilde{\Delta}_{22}} \zeta^{\epsilon}.
\end{align}
\label{seq:rescaled_equations}
\end{subequations}
where we have rescaled the noises appropriately. If we wish to set \(\tilde{D}^{\alpha\beta}= \mathds{1}_{2\times2}\), then we require that
\begin{subequations}
    \begin{align}
        \tilde{D}_{11} \cdot \epsilon^{z_{1} -2} = 1 &\implies \epsilon^{z_{1}} = \epsilon^{2}/\tilde{D}_{11}, \\
        \tilde{D}_{22} \cdot \epsilon^{z_{2} -2} = 1 &\implies \epsilon^{z_{2}} = \epsilon^{2}/\tilde{D}_{22}.
    \end{align}
\end{subequations}
This indicates that there is no way to rescale the theory to have a unit diffusion matrix without directly introducing an anisotropy in time scaling. Since the coupled SPDEs are defined with one consistent time unit, we continue with \(z_{1}=z_{2}\) and choose \(\tilde{D}_{11}\epsilon^{z_{1}-2}=1\). The noise terms can be rescaled by choosing
\begin{subequations}
    \begin{align}
    \epsilon^{z_{1}/2-\chi_{1}-1/2}\sqrt{\tilde{\Delta}_{11}} &= 1 \implies \epsilon^{\chi_{1}}=\sqrt{{\epsilon \tilde{\Delta}_{11}}/{\tilde{D}_{11}}} \\
    \epsilon^{z_{1}/2-\chi_{1}-1/2}\sqrt{\tilde{\Delta}_{11}} &= 1\implies \epsilon^{\chi_{2}} =  \sqrt{{\epsilon \tilde{\Delta}_{22}}/{\tilde{D}_{22}}}.
    \end{align}
\end{subequations}
Here, the fields are scaled anisotropically. Using this, we define new parameters based on the rescaling:
\begin{subequations}
    \begin{align}
\dfrac{1}{2}\tilde{\Gamma}^{1}_{11} \cdot \epsilon^{\chi_{1} + z_{1} -2} &= \dfrac{\tilde{\Gamma}^{1}_{11}\epsilon^{1/2}\tilde{\Delta}_{11}^{1/2}}{2\tilde{D}_{11}^{3/2}} = \sqrt{Z}\\
\dfrac{1}{2}\tilde{\Gamma}^{1}_{22} \cdot \epsilon^{2\chi_{2} - \chi_{1} + z_{1} -2} &= \dfrac{\tilde{\Gamma}^{1}_{22} \tilde{\Delta}_{22} \epsilon^{1/2} }{2\tilde{D}_{22}\tilde{D}_{11}^{1/2}\tilde{\Delta}_{11}^{1/2}} = \dfrac{\tilde{\Delta}_{22}}{\tilde{\Delta}_{11}}\dfrac{\tilde{D}_{11}}{\tilde{D}_{22}}\dfrac{\tilde{\Gamma}^{1}_{22}}{\tilde{\Gamma}^{1}_{11}}\sqrt{Z} = Y\sqrt{Z}, \\
\tilde{\Gamma}^{2}_{12} \cdot \epsilon^{\chi_{1} + z_{1} -2} &= \dfrac{\tilde{\Gamma}^{2}_{12}\epsilon^{1/2}\tilde{\Delta}_{11}^{1/2}}{\tilde{D}_{11}^{3/2}} = 2\dfrac{\tilde{\Gamma}^{2}_{12}}{\tilde{\Gamma}^{1}_{11}}\sqrt{Z} = 2X\sqrt{Z}.
    \end{align}
\end{subequations}
and, upon defining \(T=\tilde{D}_{22}/\tilde{D}_{11} \), we obtain the adimensionalised coupled KPZ equations
\begin{equation}
\begin{aligned}
\partial_{t'} \tilde{\theta}^{\epsilon}_{1} &= \partial_{x'}^{2} \tilde{\theta}^{\epsilon}_{1} +\sqrt{Z}\left( (\partial_{x'}\tilde{\theta}^{\epsilon}_{1})^{2} + Y (\partial_{x'}\tilde{\theta}^{\epsilon}_{2})^{2}\right) + \sqrt{2} \zeta^{\epsilon}, \\ 
\partial_{t'} \tilde{\theta}^{\epsilon}_{2} &= T \partial_{x'}^{2} \tilde{\theta}^{\epsilon}_{2} + 2 X\sqrt{Z}(\partial_{x'}\tilde{\theta}^{\epsilon}_{1})(\partial_{x'}\tilde{\theta}^{\epsilon}_{2}) + \sqrt{2} \zeta^{\epsilon}.
\end{aligned}
\label{eq:dedimensionalisedKardar}
\end{equation}
This shows that, unlike in the single-component case where the RG analysis involves a single parameter \(g\), the multicomponent KPZ equation requires analysis of a \(\mathbb{R}^{4}\)-space \(\{T,X,Y,Z\}\) to fully classify the theory. 
\subsection{Hyperbolicity}
\label{sec:supp-hyperbolicity}
In analysing Eq.~(\ref{eq:dedimensionalisedKardar}), we assert that Quadrants II and IV in FIG.~\ref{fig:combined}(iii) are non-hyperbolic and exhibit an instability. This instability conflicts with the scaling form leading to non-KPZ features in the long-time and long-wavelength limit and can be shown at the dynamical level. Ignoring the white noise drive and mapping to the Burger's equation for the velocity fields \(\phi'_{1}=\partial_{x}\tilde{\theta}'_{1}\) and \(\phi'_{2}=\partial_{x}\tilde{\theta}'_{2}\), we obtain 
\begin{equation}
\begin{aligned}
    \partial_{t}\vec{\phi}' = 
\begin{pmatrix}\partial_{x}^{2}&0\\0&T\partial_{x}^{2}\end{pmatrix}\vec{\phi}'+\sqrt{Z}\partial_{x}\left\{\begin{pmatrix} 
   \phi_{1} & Y\phi_{2} \\ X\phi_{2} & X\phi_{1}
    \end{pmatrix} \vec{\phi}'\right\}.
\end{aligned}
\label{eq:stability1}
\end{equation}
We seek a solution \(\phi'_{1}=\epsilon_{1} + \tilde{\phi}'_{1}(x,t)\) and \(\phi'_{2}=\epsilon_{2} + \tilde{\phi}'_{2}(x,t)\) where \((\epsilon_{1},\epsilon_{2})\in \mathbb{R}\), giving 
\begin{equation}
\begin{aligned}
    \partial_{t}\vec{\tilde{\phi}}' = 
\sqrt{Z}\begin{pmatrix} 
   2\epsilon_{1} & 2Y\epsilon_{2} \\ 2X\epsilon_{2} & 2X\epsilon_{1}
    \end{pmatrix} \vec{\tilde{\phi}}'
\end{aligned}
\end{equation}
where the Laplacian terms are sub-leading in the long-time behaviour. Stability imposes that the eigenvalues \(\lambda_{\pm} =\sqrt{Z}\left( (1+X)\epsilon_{1} \pm \sqrt{(1-X)^{2}\epsilon_{1}^{2}+4XY \epsilon_{2}^{2}}\right)\) are real for all \((\epsilon_{1}, \, \epsilon_{2})\) which is not guaranteed as the matrix is non-symmetric. In Quadrant I where \(X,\, Y > 0 \) and Quadrant III where \(X, \, Y < 0\), the discriminant is always positive. In Quadrants II and IV this is not satisfied for all \((\epsilon_{1},\,\epsilon_{2})\) leading to non-hyperbolic behaviour. We expect the RG analysis to accurately describe Quadrant I and Quadrant III where we have a scaling hypothesis. Considering compactness, the non-hyperbolicity in Quadrant IV gives rise to a proliferating STV phase with dynamical exponent \(z=1\), consistent with exponentially decaying autocorrelation function in the complex fields \(\langle\psi(t)^{\dag}\psi(t')\rangle\sim e^{-\alpha|t-t'|}\). This is explored further in Section E.
\subsection{Lyapunov and Stationary Measure}
\label{sec:supp-lyapunov}
The KPZ equation can be framed as an infinite-dimensional analogue of an It\^{o} discretised Langevin equation for a real-valued field. It is useful to understand the properties of the stationary measures, which can be motivated by considering an infinite-dimensional analogue to the Fokker-Planck equation. For fields, the probability density function gets elevated to a functional \(\mathbb{P}[\{\tilde{\theta}(x,t)\};t]\) over the dynamical fields \(\tilde{\theta}_{i}(x,t)\) at each point in space, obeying the generalisation of the Fokker-Planck equation
\begin{equation}
    \partial_{t}\mathbb{P}[\tilde{\theta}]=-\int_{x} \delta_{\tilde{\theta}^{\alpha}_{x}} \left\{\left[\tilde{D}_{\alpha\beta}\partial_{x}^{2}\tilde{\theta}^{\beta} + \tilde{\Gamma}^{\alpha}_{\beta\gamma}(\partial_{x} \tilde{\theta}^{\beta})(\partial_{x} \tilde{\theta}^{\gamma})\right]\mathbb{P}[\{\tilde{\theta}\}]\right\} + \int_{x,y} \delta_{\tilde{\theta}^{\alpha}_{x}}\delta_{\tilde{\theta}^{\alpha}_{y}}\left[\tilde{\Delta} ^{\alpha \beta} \delta(x-y) \mathbb{P}[\{\tilde{\theta}\}]\right].
\label{eq:fokkerplanckfull}
\end{equation}
The derivative is a functional derivative \(\delta_{\tilde{\theta}^{\alpha}_{x}} = \partial_{\tilde{\theta}} - \partial_{x} \cdot \partial_{\partial_{x}\tilde{\theta}^{\alpha}}\). Ignoring the KPZ non-linearity, we seek a stationary probability measure \(\partial_{t}\mathbb{P}=0\) with the Gaussian ansatz
\begin{equation}
    \mathbb{P}[\{\tilde{\theta}\}] = \int \mathcal{D}[\tilde{\theta}] \exp{- \dfrac{1}{2}\int_{x} (\partial_{x} \tilde{\theta}^{\alpha}) \tilde{C}^{-1}_{\alpha \beta} (\partial_{x} \tilde{\theta}^{\beta})}
\label{eq:stationary_measure}
\end{equation}
where \(\langle \partial_{x} \tilde{\theta}^{\alpha}\partial_{y} \tilde{\theta}^{\beta}\rangle = \tilde{C}^{\alpha\beta} \delta(x-y)\) is the symmetric same-time covariance matrix. Substituting the time stationary ansatz into Eq.~(\ref{eq:fokkerplanckfull}) gives the Lyapunov condition \(\mathbf{D} = \Delta \Sigma^{-1}\) which from symmetry gives \(\mathbf{D}\Sigma + \Sigma \mathbf{D}^{T} = 2\Delta\) relating the two-point correlation function to the diffusion and noise matrices in the non-interacting theory.  Defining rescaled interaction vertices \(\hat{\Gamma}^{\alpha} = (\tilde{C}^{-1})_{\alpha\beta} \tilde{\Gamma}^{\beta}\), analogous to the adimensionalisation scheme, the non-interacting stationary measure remains time-stationary in the presence of interactions when the cyclicity condition \(\hat{\Gamma}^{\alpha}_{\beta \gamma}=\hat{\Gamma}^{\alpha}_{\gamma\beta}=\hat{\Gamma}^{\beta}_{\gamma\alpha}\) is satisfied. The cyclicity condition for the two-component KPZ equation explicitly reads
\begin{equation}
    \hat{\Gamma}^{1} =\left(\begin{array}{cc}
    a & b \\ b & c 
    \end{array}\right), \,  \hat{\Gamma}^{2} =\left(\begin{array}{cc}
    b & c \\ c & d
    \end{array}\right).
\end{equation}
Substituting the time-stationary measure from Eq.~(\ref{eq:stationary_measure}),
\begin{equation}
\begin{aligned}
    \int dx \, \left\{\partial_{x} \left[ - \dfrac{\tilde{\Gamma}^{1}_{11}\tilde{D}^{11}}{6\tilde{\Delta}_{11}}(\partial_{x} \tilde{\theta}_{1})^{3}\right] - \left[\dfrac{\tilde{\Gamma}^{1}_{22}\tilde{D}^{11}}{2\tilde{\Delta}_{11}}(\partial_{x} \tilde{\theta}_{2})^{2} \partial_{x}^{2}\tilde{\theta}_{1} +\dfrac{\tilde{\Gamma}^{2}_{12}\tilde{D}^{22}}{\tilde{\Delta}_{22}}(\partial_{x} \tilde{\theta}_{1}) (\partial_{x} \tilde{\theta}_{2}) \partial_{x}^{2}\tilde{\theta}_{2}\right]\right\}\mathbb{P},
\end{aligned}
\label{seq:additionalstationarymeasure}
\end{equation}
and requiring that the resulting integrand is at most a total derivative results in the stationary measure condition \(\tilde{\Gamma}^{1}_{22}\tilde{D}_{11}\tilde{\Delta}_{22}=\tilde{\Gamma}^{2}_{12}\tilde{D}_{22}\tilde{\Delta}_{11}\). In this submanifold, the second term becomes a total derivative \(\partial_{x}\left[(\partial_{x}\tilde{\theta}_{2})^{2}(\partial_{x}\tilde{\theta}_{1})\right]\). This is equivalent to demanding \(X=Y\) in adimensionalised coordinates. Additionally along this submanifold, the modes are described by same-time white noise processes with static scalings coinciding with the bare non-interacting theory \(\chi_{1}=\chi_{2}=1/2\). Importantly, the RG predicts that the long-wavelength theory is described by the emergence of a stationary measure in Quadrants I and III. 
\section{C. Wilsonian Renormalisation Group}
In this section, we set up of the Wilsonian RG, starting from the MSRJD functional integral to give an equivalent action as in Eq.~(\ref{eq:kpz_path_integral}). The one-loop corrections are computed in the \(\mathbb{Z}_{2}\)-coordinates but the subsequent flows are interpreted in the adimensional coordinates in Eq.~(\ref{eq:dedimensionalisedKardar}). 
\subsection{Martin-Siggia-Rose-Janssen-De Dominicis Functional}
Motivated by Landau's symmetry arguments, we derive the general \(\mathbb{Z}_{2}\)-symmetric coupled KPZ equations, mapping to a field theory using the MSRJD formalism. The coupled equations are SPDEs driven by additive time-continuous centred Gaussian white noise processes \(\{\zeta^{\alpha}(x,t)\}_{i=1}^n\) with symmetric covariance matrix \(\Delta^{\alpha\beta} \in \mathbb{R}\) and \(\Delta^{\alpha\beta}=\Delta^{\beta\alpha}\). The noises have multivariate probability density functional \(W[\boldsymbol{\zeta}]\)
\begin{equation}
    W[\boldsymbol{\zeta}] \sim \exp{-\dfrac{1}{4} \int_{x,t} \zeta^{\alpha}\Delta_{\alpha\beta}^{-1}\zeta^{\beta}} 
    \label{eq:lidhfbjh}
\end{equation}
preserving normalisation \(\int \mathcal{D}[\boldsymbol{\zeta}] W[\boldsymbol{\zeta}] = 1\). Changing variables \(\{\boldsymbol{\zeta}(x,t)\} \mapsto \{\boldsymbol{\theta}(x,t)\}\) gives the Onsager-Machlup functional for the dynamical field probability distribution 
\begin{equation}
    \mathcal{P}[\boldsymbol{\theta}(x,t)] \sim \left|\det{\dfrac{\mathcal{D}\boldsymbol{\theta}(x,t)}{\mathcal{D}\boldsymbol{\zeta}(\boldsymbol{x}',t')}} \right|\exp{-\dfrac{1}{4} \int_{x,t}  \left(\partial_{t}\theta^{\alpha} - F^{\alpha}[\boldsymbol{\theta}]\right)\Delta_{\alpha\beta}^{-1}\left(\partial_{t}\theta^{\beta} - F^{\beta}[\boldsymbol{\theta}]\right)}
\end{equation}
Though we have taken liberty to pose the solutions to the stochastic PDE in the language of continuous functions, the solutions are often nowhere-differentiable. Even for stochastic ODEs, non-differentiability requires a discretisation scheme to properly interpret integration. This problem pervades to the level of the path integral in the Jacobian prefactor. However, the It\^o discretisation simplifies matters, as the Jacobian becomes independent of the dynamical field \(\boldsymbol{\theta}(\boldsymbol{x},t)\) and can be dropped. Introducing a purely imaginary response field \(\chi_{i} : \mathbb{R}^{d} \times \mathbb{R} \rightarrow i\mathbb{R} \)  with a Hubbard-Stratonovich transformation gives Action (\ref{eq:kpz_path_integral}). In Fourier space, using the convention in Eq.~(\ref{seq:fouriertransformconvention}) with \(k=(\omega,\vec{k})\), the action becomes
\begin{equation}
\begin{aligned}
S[\theta_{1},\theta_{2},\chi_{1},\chi_{2}] = \Bigg[\int_{k_{1}} \hspace{1mm} \Bigg[&\chi_{1,k_{1}}\left(  - i \omega \, \eta\, \theta_{1,-k_{1}} + D_{11}k^{2}_{1} \theta_{1,-k_{1}} + D_{12}k^{2}_{1} \theta_{2,-k_{1}}\right) - \chi_{1,k_{1}}\Delta_{11}\chi_{1,-k_{1}} - \chi_{1,k_{1}}\Delta_{12}\chi_{2,-k_{1}} \Bigg]\\
    + \int_{k_{1},k_{2},k_{3}}k_{2}\cdot k_{3}\Bigg[&\chi_{1,k_{1}}\left(\dfrac{\Gamma^{1}_{11}}{2} \theta_{1,k_{2}}\theta_{1,k_{3}} + \Gamma^{1}_{12} \theta_{1,k_{2}} \theta_{2,k_{3}} + \dfrac{\Gamma^{1}_{22}}{2} \theta_{2,k_{2}} \theta_{2,k_{3}}\right)\delta_{k_{1},k_{2},k_{3}}\Bigg]\Bigg] + \left(1 \leftrightarrow 2\right)
\end{aligned}
\label{eq:oawnfkn}
\end{equation}
where subscripts indicate field component and momentum. The parameter \(\eta\) scales the time derivative, but receives no RG corrections. The first line of Eq.~(\ref{eq:oawnfkn}) describes the Gaussian theory, providing the bare propagators, while the second line contains the 3-point KPZ interaction vertices with frequency and momentum conserving delta functions.  
\subsection{Non-Interacting Theory}
It is insightful to represent the Gaussian theory in Eq.~(\ref{eq:oawnfkn}) in a higher dimensional space to derive propagator rules. Integrating by parts and symmetrising, the inverse propagator matrix becomes
\begin{equation}
\begin{aligned}
    \hat{\mathbf{G}}^{-1}_{0,k_{1}} = \left(
        \scalemath{1}{\begin{array}{cccc}
        -2\Delta_{11} & i\eta\omega + D_{11} k^{2} & -2\Delta_{12} & D_{12}  k^{2} \\ -i\eta \omega + D_{11} k^{2} & 0 & D_{12}k^{2} & -2\Delta_{12} \\ -2\Delta_{12} & D_{12} k^{2} & -2\Delta_{11} & i\eta\omega + D_{11} k^{2} \\ D_{12}k^{2} & -2\Delta_{12} & -i\eta\omega + D_{11} k^{2} & 0 
        \end{array}}\right)
\end{aligned}
\label{eq:aowefjnn}
\end{equation}
for the vectorised space \(\Psi
= (\chi_{1},\theta_{1},\chi_{2},\theta_{2})\) in Fourier space. Taking functional derivatives of the generating function \(\mathcal{Z}[\boldsymbol{J}]\), the bare two-point correlators are

\begin{subequations}\label{eq:twopointfunctions}
\centering
\small{
\begin{align}
    G_{0}(k,\omega) &= \langle \theta_{\alpha,k}\chi_{\alpha,-k}\rangle =\dfrac{- D_{11}k^{2}+ i\omega}{(\omega + i (D_{11}+D_{12})k^{2})(\omega + i (D_{11}-D_{12})k^{2})} \\ 
    C_{o}(k,\omega) &=  \langle \theta_{\alpha,k}\theta_{\alpha,-k}\rangle =\dfrac{2\Delta_{11}(\omega^{2} + (D_{11}^{2}+D_{12}^{2})k^{4})-4\Delta_{12}D_{11}D_{12}k^{4}}{(\omega - i (D_{11}+D_{12})k^{2})(\omega - i (D_{11}-D_{12})k^{2})(\omega + i (D_{11}+D_{12})k^{2})(\omega + i (D_{11}-D_{12})k^{2})}\\ 
    T_{0}(k,\omega) &\underrel{\alpha \neq \beta}{=}  \langle \theta_{\alpha,k}\chi_{\beta,-k}\rangle =\dfrac{D_{12}k^{2}}{(\omega + i (D_{11}+D_{12})k^{2})(\omega + i (D_{11}-D_{12})k^{2})}\\
    Q_{0}(k,\omega) &\underrel{\alpha \neq \beta}{=}  \langle \theta_{\alpha,k}\theta_{\beta,-k}\rangle = \dfrac{2\Delta_{12}(\omega^{2} + (D_{11}^{2}+D_{12}^{2})k^{4})-4\Delta_{11}D_{11}D_{12}k^{4}}{(\omega - i (D_{11}+D_{12})k^{2})((\omega - i (D_{11}-D_{12})k^{2})(\omega + i (D_{11}+D_{12})k^{2})(\omega + i (D_{11}-D_{12})k^{2})}.
\end{align}}
\label{eq:barepropagators}
\end{subequations}
In total, there are \(16\) possible propagators, which reduce to \(8\) due to the \(\mathbb{Z}_{2}\) symmetry. The off-diagonal terms are related by inversion in frequency space \(\omega\mapsto-\omega\), from time-ordering and causality, leaving four propagators: \(G_{0},\, C_{0},\, T_{0},\, Q_{0}\). Closed response loops vanish in the It\^{o} discretisation \(\langle \chi_{\alpha,k_{1}} \chi_{\beta,-k_{1}}\rangle=0\) from causality. The response-phase connected correlation functions have poles lying in the lower-half plane, maintaining causality. 
\begin{equation}
\begin{aligned}
    \text{\includegraphics[width=0.9\textwidth]{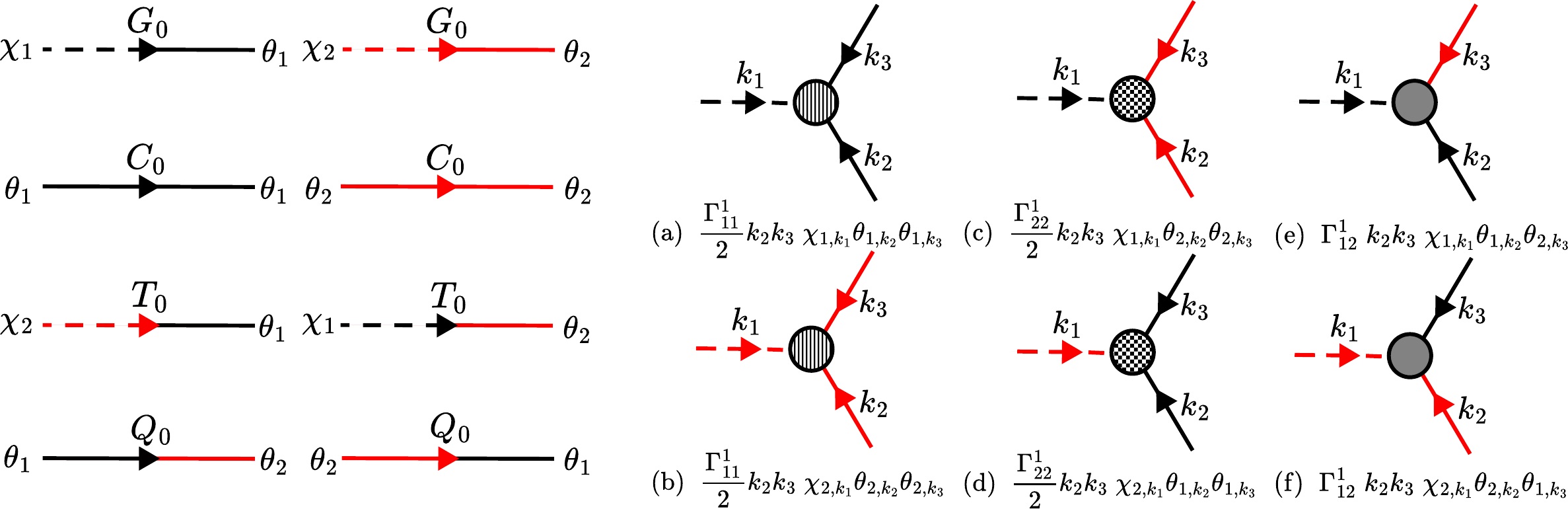}}
\end{aligned}
\end{equation}
Pictorially, response fields are represented by dashed lines with red or black edges distinguishing \((1,2)\) fields respectively. Phase fields are denoted by solid lines. Arrows signify the response field \(\chi_{\alpha,-k}\) \textit{transmutating} into a dynamical field \(\theta_{\alpha,k}\), corresponding to closing the contour in the lower-half plane for causality. Although there are \(24\) potential interaction vertices, symmetry constraints reduce this to six, with three coupling constants \(\Gamma_{11}^{1},\, \Gamma_{12}^{1},\,\Gamma_{22}^{1}\).
\subsection{Wilsonian Renormalisation Group}
Wilsonian RG generates scale-dependent theories by elevating the parameters to functions of the RG scale \(\ell\), capturing the long-wavelength physics as \(\ell \rightarrow \infty\). This can be done by iteratively integrating out fast fluctuations. The momentum is split between slow fields \(0\le k < \Lambda/b\) and fast fields \(\Lambda/b\le k \le \Lambda\), where \(\Lambda\) is the momentum cut-off regularising the theory in the UV and \(b\) is a real parameter. From linearity, \(\theta_{i} = \theta_{i}^{s}(\omega,k) + \theta_{i}^{f}(\omega,k)\). The integral measures for fast and slow fields also decouple, allowing us to define a theory over the slow modes with a statistical weight factor by integrating out the fast fields
\begin{equation}
    \mathbb{W}[\boldsymbol{\theta}^{s},\boldsymbol{\chi}^{s}] = \int D[\boldsymbol{\theta}^{f},\boldsymbol{\chi}^{f}] \exp{\underbrace{-A_{0}[\boldsymbol{\theta}^{s},\boldsymbol{\chi}^{s}]}_{\textcolor{red}{\mathrm{S\hspace{1mm}Gaussian}}}- \underbrace{A_{I}[\boldsymbol{\theta}^{s},\boldsymbol{\chi}^{s}]}_{\textcolor{red}{\mathrm{S \hspace{1mm}Int}}} - \underbrace{A_{0}[\boldsymbol{\theta}^{f},\boldsymbol{\chi}^{f}]}_{\textcolor{red}{\mathrm{F\hspace{1mm}Gaussian}}}- \underbrace{A_{I}[\boldsymbol{\theta}^{f},\boldsymbol{\chi}^{f}]}_{\textcolor{red}{\mathrm{F \hspace{1mm}Int}}}  - \underbrace{\tilde{A}_{I}[\boldsymbol{\theta}^{s},\boldsymbol{\chi}^{s},\boldsymbol{\theta}^{f},\boldsymbol{\chi}^{f}]}_{\textcolor{red}{\mathrm{S-F \hspace{1mm} Int}}}}
\label{eq:kjnaro}
\end{equation}
where \(A_{0}[\boldsymbol{\theta}^{f},\boldsymbol{\chi}^{f}]\) is the Gaussian statistical weight for fast fields. This involves computing the expectation value \(\langle\mathrm{exp}[-\tilde{A}_{I}[\boldsymbol{\theta}^{s},\boldsymbol{\chi}^{s},\boldsymbol{\theta}^{f},\boldsymbol{\chi}^{f}]]\rangle\) with respect to the probability measure over fast fields. The fast field probability distribution is typically assumed to be Gaussian, as including interaction terms leads to higher-order corrections than the \(1\)-loop. Thus the effective statistical weight for the slow fields becomes:
\begin{equation}
   \mathbb{W}[\boldsymbol{\theta}^{s},\boldsymbol{\chi}^{s}] =  e^{-A[\boldsymbol{\theta}^{s},\boldsymbol{\chi}^{s}]} \langle e^{-\tilde{A}_{I}[\boldsymbol{\theta}^{s},\boldsymbol{\chi}^{s},\boldsymbol{\theta}^{f},\boldsymbol{\chi}^{f}]}\rangle_{f}
\end{equation}
where \(\langle \cdot \rangle_{f}\) indicates integration over fast fields using Wick's theorem such that all cumulants vanish for \(n>2\). The Wilsonian Effective Action is defined as: 
\begin{equation}
A'[\boldsymbol{\theta}^{s},\boldsymbol{\chi}^{s}]=A[\boldsymbol{\theta}^{s},\boldsymbol{\chi}^{s}] - \log{\langle e^{-\tilde{A}_{I}[\boldsymbol{\theta}^{s},\boldsymbol{\chi}^{s},\boldsymbol{\theta}^{f},\boldsymbol{\chi}^{f}]}\rangle_{f}}
\label{seq:wilsonianeffectiveaction}
\end{equation}
generating the connected subset of all possible graphs. To calculate the renormalised parameters, we use a perturbative expansion in interactions around the Gaussian fixed point. Expanding the logarithm yields the corrections: 
\begin{equation}
    \mathrm{Corrections} = \langle\tilde{A}_{I}\rangle_{c} -\dfrac{1}{2!} \langle \tilde{A}_{I}^{2}\rangle_{c} + \dfrac{1}{3!} \langle \tilde{A}_{I}^{3}\rangle_{c}.
\label{eq:knjasdjkb}
\end{equation}
Calculating the renormalised terms to \(1\)-loop requires automation of generating connected diagrams. An Important point to highlight is that integration over \(\chi_{i}^{f}\) is over the imaginary axis from \((-i\infty, i \infty)\) yielding an additional minus sign from Wick rotation. In the \(1\)-loop corrections, we assume analyticity of the resulting integrals in slow momenta \(k_{s}\) and frequency \(\omega_{s}\). The corrections involve integrals of the form:
\begin{equation}
\begin{aligned}
    \int_{\omega,k} \hspace{1mm} f(\omega, k,\omega_{s},k_{s}) &= \int_{\omega,k} \sum_{n,m=0} \omega_{s}^{n} k_{s}^{m} f^{(n,m)}(\omega, k)\\
    & \stackrel{!}{=} \sum_{n,m=0} \omega_{s}^{n} k_{s}^{m}\int_{\omega,k}f^{(n,m)}(\omega, k)
\end{aligned}
\label{eq:dsjflhbf}
\end{equation}
where \(f\) is a product of propagators and interaction vertices. \((k,\omega)\) are the free-momentum around the loop and \((k_{s},\omega_{s})\) are the incoming momentum. If the order of the infinite sum and integral can be exchanged, the relevant corrections can be found in the series expansion in slow variables. In the spirit of non-covariant theories, there is no cutoff in frequency, and the resulting frequency integral runs over the real line. The domain of integration for the momentum in Cartesian coordinates is given by the set of points for which \(\mathcal{D} = \{(x_{1},x_{2},\ldots) | \Lambda/b \le \sqrt{\sum x_{i}^{2}} \le \Lambda\}\) defining a shell in momentum space for the fast modes. Two frequently appearing integrals are:
\begin{equation}
\begin{aligned}
    \int_{\mathcal{D}} \dfrac{d^{d}q}{(2\pi)^{d}} \dfrac{1}{q^{2}} 
    =  \dfrac{1}{2^{d-1}\Gamma(d/2)\pi^{d/2}}\Lambda^{d-2} \;d\ell 
    = K[d] \;d\ell 
\end{aligned}
\end{equation}
which diverges in the infrared as \(q\rightarrow 0\) for \(d_{c}\le 2\). This divergence does not appear in our integrals over a slice, and
\begin{equation}
\begin{aligned}
    \int_{\mathcal{D}}\dfrac{d^{d}q}{(2\pi)^{d}}  \dfrac{(k_{s}\cdot q)^{2}}{(q^{4})} &= p^{i}_{s} p^{j}_{s}\int_{\mathcal{D}} \dfrac{d^{d}q}{(2\pi)^{d}}  \dfrac{q^{i}q^{j}} {(q^{4})} != p^{i}_{s} p^{j}_{s}\dfrac{\delta_{ij}}{d}\int_{\mathcal{D}} \dfrac{d^{d}q}{(2\pi)^{d}} \dfrac{1}{q^{2}} = \dfrac{k_{s}\cdot k_{s}}{d}\; k[d] \;d\ell.
\end{aligned}
\end{equation}
which appeals to Passerino-Veltman arguments. At this stage, we calculate the corrections from Eq.~(\ref{eq:knjasdjkb}) for the parameters in the \(\mathbb{R}^{8}\)-parameter space \(\mathcal{P}=\{\eta, D_{11},D_{12}, \Gamma^{1}_{11},\Gamma^{1}_{12}, \Gamma^{1}_{22},\Delta_{11},\Delta_{12}\}\). No corrections are generated for \(\eta\) which requires the generation of \(\omega_{s}\chi^{\alpha}_{s}\theta^{\alpha}_{s}\) in the Wilsonian effective action. The non-covariant theory exhibits anisotropic scaling in space and time 
\begin{equation}
    x\mapsto bx \hspace{1mm},\hspace{1mm} t \mapsto b^{z}t\hspace{1mm},\hspace{1mm} \theta^{\alpha}\mapsto b^{\chi}\theta^{\alpha}\hspace{1mm},\hspace{1mm} \chi^{\alpha}\mapsto b^{\tilde{\xi}}\chi^{\alpha} 
\end{equation}
as well as scaling to the response fields. We assume a single time-scaling parameter, which is valid in the non-decoupled case of Eq.~(\ref{eq:dedimensionalisedKardar}) within the normal mode space. However, \(X=0\) is an exceptional point allowing independent time rescaling for both modes. This is equivalent to demanding that in the RG limit, there is one scaling parameter and the resulting two-point functions take on the scaling form 
\begin{equation}
    \langle \theta_{\alpha}\theta_{\beta}\rangle \sim t^{\chi/z} C_{\alpha\beta}(x/t^{1/z})
\end{equation}
where \(\chi\) and \(z\) are to be determined. The \(\mathbb{Z}_{2}\) symmetry is preserved under RG. The corrections for each parameter are calculated, and comparisons are drawn to the single-component case.
\subsection{Corrections to Covariance Matrix}
The one-loop corrections to the covariance matrix \(\Delta_{\alpha\beta}\) arise from terms in the expansion of the Wilsonian effective action (\ref{seq:wilsonianeffectiveaction}) proportional to \(\chi^{s}_{\alpha}\chi^{s}_{\beta}\), specifically in \(-(1/2)\langle\tilde{A}_{I}^{2}\rangle_{c}\). For instance, corrections for \(\Delta_{11}\) can be represented by diagrammatic sums
\begin{equation}
\begin{aligned}
    \text{\includegraphics[width=0.95\textwidth]{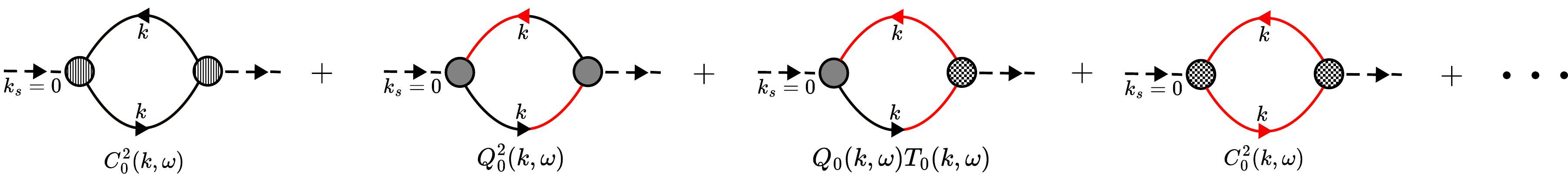}}
\end{aligned}
\end{equation}
where \(\chi^{s}_1\) are outgoing-legs. The internal propagators are products of phase-phase correlators, either \(Q_{0}\) or \(C_{0}\). The degeneracy of each diagram is calculated via automated combinatorics by directly expanding the action. The corrections to \(\Delta_{11}\) are simplified by setting in-going momentum to zero from argument Eq.~(\ref{eq:dsjflhbf}). The loop corrections, \(\loopdiagram\), for \(\Delta_{11}\) and \(\Delta_{12}\) are: 
\begin{equation}
\begin{gathered}
     \left[\loopdiagram\right]_{\Delta_{11}}=-\dfrac{1}{4} \int \dfrac{d^{d}k}{(2\pi)^{d}}\dfrac{d\omega}{2\pi} (k\cdot k)^{2}\left[((\Gamma^{1}_{11})^{2} + 2(\Gamma^{1}_{12})^{2}+(\Gamma^{1}_{22})^{2})C_{0}^{2} + 4 \Gamma^{1}_{12}(\Gamma^{1}_{11}+\Gamma^{1}_{22})C_{0}Q_{0} + 2((\Gamma^{1}_{12})^{2} + \Gamma^{1}_{11}\Gamma^{1}_{22})Q_{0}^{2}\right]\\
      \left[\loopdiagram\right]_{\Delta_{12}}=-\dfrac{1}{4} \int \dfrac{d^{d}k}{(2\pi)^{d}}\dfrac{d\omega}{2\pi} (k\cdot k)^{2}\left[((\Gamma^{1}_{11})^{2} + 2(\Gamma^{1}_{12})^{2}+(\Gamma^{1}_{22})^{2})Q_{0}^{2} + 4 \Gamma^{1}_{12}(\Gamma^{1}_{11}+\Gamma^{1}_{22})C_{0}Q_{0} + 2((\Gamma^{1}_{12})^{2} + \Gamma^{1}_{11}\Gamma^{1}_{22})C_{0}^{2}\right]
\end{gathered}
\end{equation}
where the \(1/4\) prefactor comes from symmetrising the corrections \(\Delta_{12}=\Delta_{21}\) and the minus comes from the expansion. By \(\mathbb{Z}_{2}\) symmetry, \(\Delta_{11}\)=\(\Delta_{22}\) and the flow equation can be deduced directly. The integrals over frequency are calculated by closing the integral in the lower complex plane and using Jordan's Lemma and Cauchy Residue Theorem. The covariance matrix corrections are
\begin{equation}
\begin{gathered}
    -  \int_{x, t} \left[\Delta_{\alpha \beta}b^{z+2\tilde{\xi} + d}- \loopdiagram\right] \chi_{\alpha}\chi_{\beta}\; \implies \; \partial_{\ell} \Delta_{\alpha \beta} = \left(z+ 2\tilde{\xi} + d -  \dfrac{\left[\loopdiagram\right]_{\Delta_{\alpha\beta}}}{\Delta_{\alpha\beta}}\right)\Delta_{\alpha\beta}
\end{gathered}
\end{equation}
from which the infinitesimal RG equations are calculated. They become more instructive in the normal mode basis where: 
\begin{equation}
\begin{gathered}
    \left[\loopdiagram\right]_{\tilde{\Delta}_{11}} = -\frac{K[d] \left((\tilde{\Gamma}^{1}_{11})^2 \tilde{\Delta}_{11}^2 \tilde{D}_{22}^3+\tilde{D}_{11}^3 (\tilde{\Gamma}^{1}_{22})^2 \tilde{\Delta}_{22}^2\right)}{4\tilde{D}_{11}^3 \tilde{D}_{22}^3}, \;
    \left[\loopdiagram\right]_{\tilde{\Delta}_{22}} = -\frac{K[d] \left((\tilde{\Gamma}^{2}_{12})^2 \tilde{\Delta}_{11}\tilde{\Delta}_{22}\right)}{\tilde{D}_{11}\tilde{D}_{22}(\tilde{D}_{11}+\tilde{D}_{22})}.
\end{gathered}
\label{seq:correctioncovariance}
\end{equation}
As a consistency check, the corrections to \(\tilde{\Delta}_{11}\) agree with the single-component KPZ RG corrections from \(\tilde{\Gamma}_{11}\), but with an additional contribution from \(\tilde{\Gamma}^{1}_{22}\) with the same structure but different propagators. 
\subsection{Corrections to Diffusion Matrix}
For the \(1\)-loop corrections to the diffusive matrix elements \(D_{11}\) and \(D_{12}\), we introduce non-zero incoming and outgoing momenta, labelled as \(k_{s}\) to generate terms of the form \(k_{s}^{2} \chi^{s}_{\alpha}\theta^{s}_{\beta}\). The corrections to \(D_{11}\) are
\begin{equation}
\begin{aligned}
\text{\includegraphics[width=0.95\textwidth]{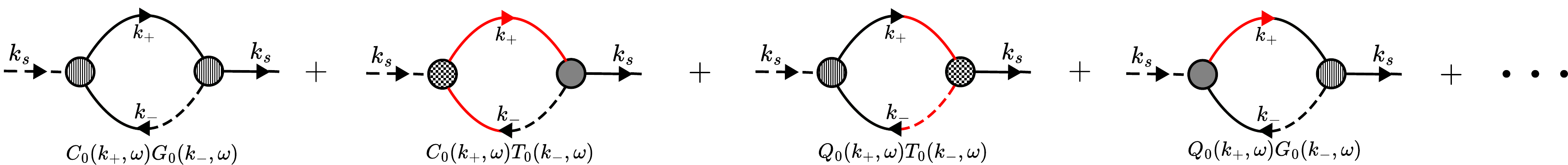}}
\end{aligned}
\label{seq:correction_d11}
\end{equation}
where \(k_{s}\) is the incoming momentum and \(k_{\pm}= k \pm \frac{k_{s}}{2}\). Since there is no outgoing slow frequency dependence, \(\omega_{s}=0\) ensuring the frequency in the internal loop is the same. The momentum dependence also appears in the vertex factors. All diagrams in Eq.~(\ref{seq:correction_d11}) have the same topology with momentum contribution factor
\begin{equation}
\begin{aligned}
f(k,k_{s}) &=\left[k+k_{s}/2\right]\cdot\left[k-k_{s}/2\right] \left[k+k_{s}/2\right]\cdot\left[ -k_{s}\right]
\end{aligned}
\end{equation}
where the interaction vertex has convention that the momentums are incoming. Focusing on terms proportional to \(k_{s}\cdot k_{s}\), the total correction \(\loopdiagram\) is given by:
\begin{equation}
\begin{aligned}
    \left[\loopdiagram\right]_{D_{11}}=-\int \dfrac{d^{d}k}{(2\pi)^{d}}\dfrac{d\omega}{2\pi} f(k,k_{s})\Big[&((\Gamma^{1}_{11})^{2}+(\Gamma^{1}_{12})^{2} + 2\Gamma^{1}_{12}\Gamma^{1}_{22})C_{0,k_{+}}G_{0,k_{-}} + ((\Gamma^{1}_{12})^{2} + 2\Gamma^{1}_{11}\Gamma^{1}_{12} + (\Gamma^{1}_{22})^{2})Q_{0,k_{+}}G_{0,k_{-}}\\ 
    &+ (\Gamma^{1}_{11} + \Gamma^{1}_{12})(\Gamma^{1}_{22} + \Gamma^{1}_{12})\left(C_{0,k_{+}}T_{0,k_{-}} + Q_{0,k_{+}}T_{0,k_{-}}\right)
    \Big]
\end{aligned}
\end{equation}
and for the off-diagonal term \(D_{12}\)
\begin{equation}
\begin{aligned}
    \left[\loopdiagram\right]_{D_{12}}=-\int \dfrac{d^{d}k}{(2\pi)^{d}}\dfrac{d\omega}{2\pi} f(k,k_{s})\Big[&((\Gamma^{1}_{11})^{2}+(\Gamma^{1}_{12})^{2} + 2\Gamma^{1}_{12}\Gamma^{1}_{22})Q_{0,k_{+}}T_{0,k_{-}} + ((\Gamma^{1}_{12})^{2} + 2\Gamma^{1}_{11}\Gamma^{1}_{12} + (\Gamma^{1}_{22})^{2})C_{0,k_{+}}T_{0,k_{-}}\\ 
    &+ (\Gamma^{1}_{11} + \Gamma^{1}_{12})(\Gamma^{1}_{22} + \Gamma^{1}_{12})\left(Q_{0,k_{+}}G_{0,k_{-}} + C_{0,k_{+}}G_{0,k_{-}}\right)
    \Big].
\end{aligned}
\end{equation}
We first integrate over \(\omega\) to zeroth order in \(\omega_{s}\), then expand the integrand to \(O(k_{s}^{2})\). The action receives correction
\begin{equation}
\begin{gathered}
    - \int_{\omega, k} \left[D_{\alpha \beta} b^{z+\chi + \tilde{\xi} + d-2} +  \loopdiagram\right] \chi_{\alpha}\grad^{2} \theta_{\beta} \; \implies \; \partial_{\ell} D_{\alpha \beta} = \left(z+ \chi + \tilde{\xi} + d - 2 +  \dfrac{\left[\loopdiagram\right]_{D_{\alpha\beta}}}{D_{\alpha\beta}}\right)D_{\alpha\beta}
\end{gathered}
\end{equation}
where the additional minus comes from going back to the real space representation of the fields \(q_{s}^{2}\mapsto -\grad^{2}\). The corrections in the rotated basis become: 
\begin{equation}
\begin{gathered}
    \scalebox{1}{$
    \left[\loopdiagram\right]_{\tilde{D}_{11}} = \dfrac{\left(2-d\right)\left(\tilde{D}_{22}^{2}\tilde{\Delta}_{11}(\tilde{\Gamma}^{1}_{11})^{2} + \tilde{D}_{11}^{2}\tilde{\Delta}_{22}\tilde{\Gamma}^{1}_{22}\tilde{\Gamma}^{2}_{12}\right)}{4d\tilde{D_{11}}^{2}\tilde{D_{22}}^{2}}K[d]$}\\
   \scalebox{1}{$\left[\loopdiagram\right]_{\tilde{D}_{22}} = \left(\tilde{\Gamma}^{1}_{22} \tilde{\Gamma}^{2}_{12} \tilde{\Delta}_{22} M_{\tilde{D}_{11},\tilde{D}_{22}} + (\tilde{\Gamma}^{2}_{12})^{2}\tilde{\Delta}_{11}M_{\tilde{D}_{22},\tilde{D}_{11}}\right) K[d]$}\\
   \scalebox{1}{$ M_{\tilde{D}_{11},\tilde{D}_{22}}= \dfrac{\left((4-d) \tilde{D}_{22}- d \tilde{D}_{11}\right)}{2 d \tilde{D}_{22} \left(\tilde{D}_{11}+\tilde{D}_{22}\right){}^2}$}.
\end{gathered}
\label{seq:correction_diffusion}
\end{equation}
For \(\tilde{D}_{11}\), there is the standard KPZ correction from the single-component and an additional cross-component with the same structure. In agreement with the single-component KPZ equation, the corrections to \(\tilde{D}_{11}\) vanish at \(d_{c}=2\). The term \(\tilde{D}_{22}\) is renormalised only by the cross-component interaction, resulting in a different structure. 
\subsection{Corrections to Non-linear Terms}
In the single-component case, the non-linear term does not receive corrections due to the infinitesimal Galilean invariance of the MSR action, which leads to Ward-Takahashi Identities relating the RG flow of \(\eta\) to the non-linearity \(\Gamma\) \cite{PhysRevE.84.061128}. In the coupled \(\mathbb{Z}_{2}\) theory, this symmetry is lost and all \(\Gamma^{\alpha}_{\beta\gamma}\) receive corrections. There are three distinct diagrammatic topologies based on momentum flow: (a1-a2) asymmetric graphs and (s1) symmetric graphs. Depending on the momentum flow and the direction of the retarded Green's functions, each topology has a different number of poles in the lower half-plane. To illustrate this, we examine the intracomponent contributions to \(\Gamma^{1}_{11}\), where no propagators contain fields indexed by 2, as shown below:  
\begin{equation}
\begin{aligned}
\text{\includegraphics[width=0.9\textwidth]{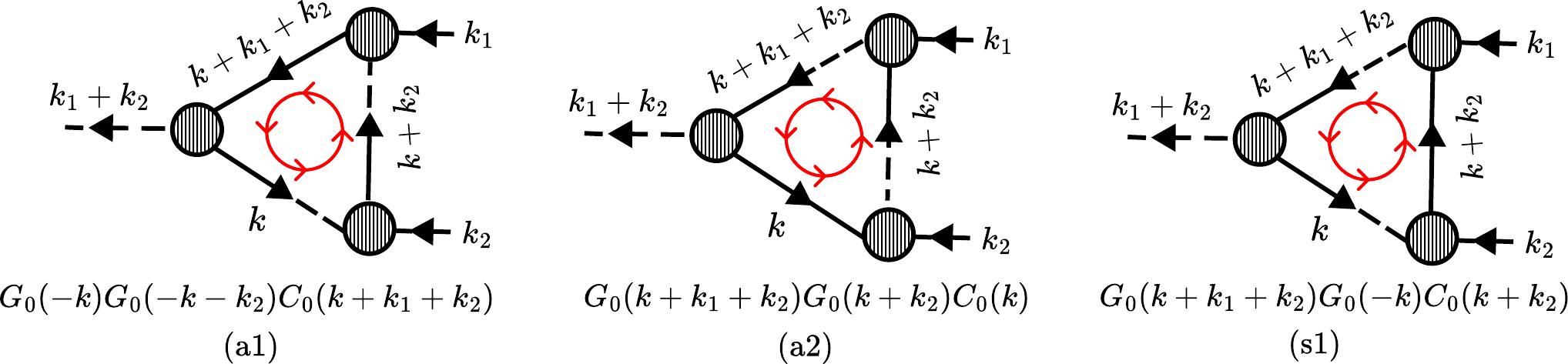}}.
\end{aligned}
\label{seq:non_linear}
\end{equation}
In these diagrams, \(k_{1}\) and \(k_{2}\) are the incoming momenta of the phase fields and the red loop indicates the internal momentum flow. Corrections from the asymmetric terms (a1-a2) are equal but differ in the number of poles in the lower half-plane: (a1) has two poles, (a2) has six poles, and (s1) has four poles. Moreover, the momentum dependent vertex factors are different between topologies. We include the full expressions for the corrections which are long and sadly not that enlightening. The symmetric contributions  (s1) are:
\begin{equation}
\begin{aligned}
&\left[\loopdiagram\right]_{\Gamma^{1}_{11}}^{\mathrm{s1}}=-\int \dfrac{d^{d}k}{(2\pi)^{d}}\dfrac{d\omega}{2\pi} f_{(\mathrm{s1})}(k,k_{1},k_{2})\Big[M_{1}\left(G_{0,-k}G_{0,q}C_{0,r}\right)+ M_{2}\left(T_{0,-k}G_{0,q}C_{0,r} + G_{0,-k}T_{0,q}C_{0,r}\right)\\&+ M_{3}\left(T_{0,-k}T_{0,q}C_{0,r}\right) + M_{4}\left(G_{0,-k}G_{0,q}Q_{0,r}\right)+  M_{5}\left(T_{0,-k}G_{0,q}Q_{0,r} +  G_{0,-k}T_{0,q}Q_{0,r}\right)+ M_{6}\left(T_{0,-k}T_{0,q}Q_{0,r}\right)\Big],\\
&\left[\loopdiagram\right]_{\Gamma^{1}_{12}}^{\mathrm{s1}}=-\int \dfrac{d^{d}k}{(2\pi)^{d}}\dfrac{d\omega}{2\pi} f_{(\mathrm{s1})}(k,k_{1},k_{2})\Big[M_{5}\left(G_{0,-k}G_{0,q}C_{0,r}\right)+ M_{4}\left(T_{0,-k}G_{0,q}C_{0,r}\right) + M_{6}\left(G_{0,-k}T_{0,q}C_{0,r}\right)\\&+ M_{5}\left(T_{0,-k}T_{0,q}C_{0,r}\right) + M_{2}\left(G_{0,-k}G_{0,q}Q_{0,r}\right)+  M_{1}\left(T_{0,-k}G_{0,q}Q_{0,r}\right) +  M_{3}\left(G_{0,-k}T_{0,q}Q_{0,r}\right)+ M_{2}\left(T_{0,-k}T_{0,q}Q_{0,r}\right)\Big],\\
&\left[\loopdiagram\right]_{\Gamma^{1}_{22}}^{\mathrm{s1}}=-\int \dfrac{d^{d}k}{(2\pi)^{d}}\dfrac{d\omega}{2\pi} f_{(\mathrm{s1})}(k,k_{1},k_{2})\Big[M_{3}\left(G_{0,-k}G_{0,q}C_{0,r}\right)+ M_{2}\left(T_{0,-k}G_{0,q}C_{0,r} + G_{0,-k}T_{0,q}C_{0,r}\right)\\&+ M_{1}\left(T_{0,-k}T_{0,q}C_{0,r}\right) + M_{6}\left(G_{0,-k}G_{0,q}Q_{0,r}\right)+  M_{5}\left(T_{0,-k}G_{0,q}Q_{0,r} +  G_{0,-k}T_{0,q}Q_{0,r}\right)+ M_{4}\left(T_{0,-k}T_{0,q}Q_{0,r}\right)\Big],
\end{aligned}
\end{equation}
where \(q= k+k_{1}+k_{2}\) and \(r=k+k_{2}\) and \(M_{i}\) are constants that are polynomials of interaction vertices
\begin{equation}
\begin{aligned}
    M_{1} & = -\frac{1}{2} \left((\Gamma^{1}_{11})^{3}+\Gamma^{1}_{12} \left(\Gamma^{1}_{12}+2 \Gamma^{1}_{22}\right) \Gamma^{1}_{11}+2 (\Gamma^{1}_{12})^{3}+(\Gamma^{1}_{22})^{3}-(\Gamma^{1}_{12})^{2}
   \Gamma^{1}_{22}\right)\\
   M_{2} &= - \frac{1}{2} \left(\left(\left(\Gamma^{1}_{12}+\Gamma^{1}_{22}\right) (\Gamma^{1}_{11})^{2}\right)+\left((\Gamma^{1}_{12})^{2}+(\Gamma^{1}_{22})^{2}\right) \Gamma^{1}_{11}+\Gamma^{1}_{12} \left(2 (\Gamma^{1}_{12})^{2}+\Gamma^{1}_{22} \Gamma^{1}_{12}+(\Gamma^{1}_{22})^{2}\right)\right)\\
   M_{3} &= -\frac{1}{2} \left(\Gamma^{1}_{11}+2 \Gamma^{1}_{12}+\Gamma^{1}_{22}\right) \left((\Gamma^{1}_{12})^2+\Gamma^{1}_{11} \Gamma^{1}_{22}\right)\\
   M_{4} &= -\Gamma^{1}_{12} \left((\Gamma^{1}_{11})^{2}+\Gamma^{1}_{12} \Gamma^{1}_{11}+\Gamma^{1}_{22}
   \left(\Gamma^{1}_{12}+\Gamma^{1}_{22}\right)\right)\\
   M_{5} &= -\frac{1}{2} \Gamma^{1}_{12} \left(\Gamma^{1}_{11}+\Gamma^{1}_{22}\right) \left(\Gamma^{1}_{11}+2 \Gamma^{1}_{12}+\Gamma^{1}_{22}\right)\\ 
   M_{6} &= -\Gamma^{1}_{12} \left(\Gamma^{1}_{12} \Gamma^{1}_{22}+\Gamma^{1}_{11} \left(\Gamma^{1}_{12}+2
   \Gamma^{1}_{22}\right)\right).
\end{aligned}
\label{seq:vertexfunctionscorrectionsgammasymmetric}
\end{equation}
The asymmetric terms (a1-a2) are equal. The corrections to the (a1) topologies are: 
\begin{equation}
\begin{aligned}
&\left[\loopdiagram\right]_{\Gamma^{1}_{11}}^{\mathrm{a1}}=-\int \dfrac{d^{d}k}{(2\pi)^{d}}\dfrac{d\omega}{2\pi} f_{(\mathrm{a1})}(k,k_{1},k_{2})\Big[\tilde{M}_{1}\left(G_{0,-k}G_{0,-r}C_{0,q}\right)+ \tilde{M}_{2}\left(T_{0,-k}G_{0,-r}C_{0,q}\right) + \tilde{M}_{3}\left(G_{0,-k}T_{0,-r}C_{0,q}\right)\\&+ \tilde{M}_{4}\left(T_{0,-k}T_{0,-r}C_{0,q}\right) + \tilde{M}_{5}\left(G_{0,-k}G_{0,-r}Q_{0,q}\right)+  \tilde{M}_{6}\left(T_{0,-k}G_{0,-r}Q_{0,q}\right) + \tilde{M}_{7}\left(G_{0,-k}T_{0,-r}Q_{0,q}\right)+ \tilde{M}_{8}\left(T_{0,-k}T_{0,-r}Q_{0,q}\right)\Big],\\
&\left[\loopdiagram\right]_{\Gamma^{1}_{12}}^{\mathrm{a1}}=-\int \dfrac{d^{d}k}{(2\pi)^{d}}\dfrac{d\omega}{2\pi} f_{(\mathrm{a1})}(k,k_{1},k_{2})\Big[\tilde{M}_{4}\left(G_{0,-k}G_{0,-r}C_{0,q}\right)+ \tilde{M}_{3}\left(T_{0,-k}G_{0,-r}C_{0,q}\right) + \tilde{M}_{2}\left(G_{0,-k}T_{0,-r}C_{0,q}\right)\\&+ \tilde{M}_{1}\left(T_{0,-k}T_{0,-r}C_{0,q}\right) + \tilde{M}_{8}\left(G_{0,-k}G_{0,-r}Q_{0,q}\right)+  \tilde{M}_{7}\left(T_{0,-k}G_{0,-r}Q_{0,q}\right) + \tilde{M}_{6}\left(G_{0,-k}T_{0,-r}Q_{0,q}\right)+ \tilde{M}_{5}\left(T_{0,-k}T_{0,-r}Q_{0,q}\right)\Big],\\
&\left[\loopdiagram\right]_{\Gamma^{1}_{22}}^{\mathrm{a1}}=-\int \dfrac{d^{d}k}{(2\pi)^{d}}\dfrac{d\omega}{2\pi} f_{(\mathrm{a1})}(k,k_{1},k_{2})\Big[\tilde{M}_{6}\left(G_{0,-k}G_{0,-r}C_{0,q}\right)+ \tilde{M}_{5}\left(T_{0,-k}G_{0,-r}C_{0,q}\right) + \tilde{M}_{8}\left(G_{0,-k}T_{0,-r}C_{0,q}\right)\\&+ \tilde{M}_{7}\left(T_{0,-k}T_{0,-r}C_{0,q}\right) + \tilde{M}_{2}\left(G_{0,-k}G_{0,-r}Q_{0,q}\right)+  \tilde{M}_{1}\left(T_{0,-k}G_{0,-r}Q_{0,q}\right) + \tilde{M}_{4}\left(G_{0,-k}T_{0,-r}Q_{0,q}\right)+ \tilde{M}_{3}\left(T_{0,-k}T_{0,-r}Q_{0,q}\right)\Big],
\end{aligned}
\end{equation}
where the corrections involve sums of products of propagators, with different prefactors depending on the allowed interaction vertices. The constants \(\tilde{M}_{i}\), distinct from the symmetric graph corrections Eq.~(\ref{seq:vertexfunctionscorrectionsgammasymmetric}), are 
\begin{equation}
\begin{aligned}
    \tilde{M}_{1} & = -\frac{1}{2} \left((\Gamma^{1}_{11})^{3}+\Gamma^{1}_{12} \left(\Gamma^{1}_{12}+2 \Gamma^{1}_{22}\right) \Gamma^{1}_{11}+\Gamma^{1}_{12} \left(\Gamma^{1}_{12}+\Gamma^{1}_{22}\right){}^2\right)\\
   \tilde{M}_{2} &= - \frac{1}{2} \left((\Gamma^{1}_{12})^{3}+(\Gamma^{1}_{11})^{2} \Gamma^{1}_{12}+\left((\Gamma^{1}_{11})^{2}+2 \Gamma^{1}_{12}
   \Gamma^{1}_{11}+3 (\Gamma^{1}_{12})^{2}\right) \Gamma^{1}_{22}\right)\\
   \tilde{M}_{3} &= -\frac{1}{2} \left(\Gamma^{1}_{12} \left(\Gamma^{1}_{11}+\Gamma^{1}_{12}\right){}^2+2 \Gamma^{1}_{12} (\Gamma^{1}_{22})^{2}+\left((\Gamma^{1}_{11})^{2}+(\Gamma^{1}_{12})^{2}\right) \Gamma^{1}_{22}\right)\\
   \tilde{M}_{4} &= -\frac{1}{2} \left((\Gamma^{1}_{12})^{3}+2 \Gamma^{1}_{22} (\Gamma^{1}_{12})^{2}+(\Gamma^{1}_{11})^{2} \Gamma^{1}_{12}+\Gamma^{1}_{11} \left(\Gamma^{1}_{12}+\Gamma^{1}_{22}\right){}^2\right)\\
   \tilde{M}_{5} &= - \dfrac{1}{2}\left(2 \Gamma^{1}_{12} (\Gamma^{1}_{11})^{2}+\left((\Gamma^{1}_{12})^{2}+(\Gamma^{1}_{22})^{2}\right) \Gamma^{1}_{11}+\Gamma^{1}_{12} \left(\Gamma^{1}_{12}+\Gamma^{1}_{22}\right){}^2\right)\\
    \tilde{M}_{6} &= -\dfrac{1}{2}\left((\Gamma^{1}_{12})^{3}+2 \Gamma^{1}_{11} (\Gamma^{1}_{12})^{2}+(\Gamma^{1}_{22})^{2} \Gamma^{1}_{12}+\left(\Gamma^{1}_{11}+\Gamma^{1}_{12}\right){}^2 \Gamma^{1}_{22}\right)\\
    \tilde{M}_{7} &= -\dfrac{1}{2}\left((\Gamma^{1}_{12})^{3}+2 \Gamma^{1}_{11} (\Gamma^{1}_{12})^{2}+(\Gamma^{1}_{22})^{2} \Gamma^{1}_{12}+\left(\Gamma^{1}_{11}+\Gamma^{1}_{12}\right){}^2 \Gamma^{1}_{22}\right)\\
    \tilde{M}_{8} &= -\dfrac{1}{2}\left((\Gamma^{1}_{12})^{3}+(\Gamma^{1}_{22})^{2} \Gamma^{1}_{12}+\Gamma^{1}_{11} \left(3 (\Gamma^{1}_{12})^{2}+2
   \Gamma^{1}_{22} \Gamma^{1}_{12}+(\Gamma^{1}_{22})^{2}\right)\right)
\end{aligned}
\end{equation}
There are 8 distinct products of propagators. Expanding to zeroth order in \(\omega_{s}\), the frequency integral is computed using Cauchy residue theorem and expanded in powers of \(k_{1}, \, k_{2}\). The MSR action transforms as:
\begin{equation}
\begin{gathered}
    - \int_{\omega, k} \left[ \dfrac{\Gamma^{\alpha}_{\beta \gamma}}{2}b^{z+2\chi + \tilde{\xi} + d-2} +  \loopdiagram\right] \chi_{\alpha}(\grad\theta_{\beta})(\grad\theta_{\gamma}) \; \implies \; \partial_{\ell} \Gamma^{\alpha}_{\beta \gamma} = \left(z+ 2\chi + \tilde{\xi} + d - 2 +  \dfrac{\left[\loopdiagram\right]_{\Gamma^{\alpha}_{\beta \gamma}}}{\Gamma^{\alpha}_{\beta \gamma}}\right)\Gamma^{\alpha}_{\beta \gamma}.
\end{gathered}
\label{seq:flows_gamma}
\end{equation}
In the normal mode basis, the non-linear corrections simplify with \(\tilde{\Gamma}^{1}_{11}\) vanishing at one-loop and the other two corrections: 
\begin{equation}
\begin{aligned}
\left[\loopdiagram\right]_{\tilde{\Gamma}^{1}_{22}} = K[d]\frac{\tilde{\Gamma}^{1}_{22} \left(\tilde{\Gamma}^{1}_{11}  \tilde{D}_{22}-\tilde{D}_{11} \tilde{\Gamma}^{2}_{12}\right) \left(\tilde{D}_{11} \tilde{\Gamma}^{1}_{22} \tilde{\Delta}_{22}-\tilde{\Delta}_{11} \tilde{D}_{22} \tilde{\Gamma}^{2}_{12}\right)}{d \tilde{D}_{11}^2 \tilde{D}_{22}^2 \left(\tilde{D}_{11}+\tilde{D}_{22}\right)}\\     \left[\loopdiagram\right]_{\tilde{\Gamma}^{2}_{12}} =  K[d]\frac{\left(\tilde{\Gamma}^{1}_{11} -\tilde{\Gamma}^{2}_{12}\right) \tilde{\Gamma}^{2}_{12} \left(\tilde{\Delta}_{11} \tilde{D}_{22} \tilde{\Gamma}^{2}_{12}-\tilde{D}_{11} \tilde{\Gamma}^{1}_{22} \tilde{\Delta}_{22}\right)}{d \tilde{D}_{11} \tilde{D}_{22} \left(\tilde{D}_{11}+\tilde{D}_{22}\right){}^2}
\end{aligned}
\label{seq:correctiongamma}
\end{equation}
The structure of these corrections suggest the existence of fixed points, such as the fixed line \(X=Y\) where both non-linearities receive no correction. This corresponds to the stable fixed points in Quadrants I and III and the emergence of a fluctuation-dissipation relation and stationary measure.
\subsection{Flow Equations}
Up to this point, the corrections Eq.~(\ref{seq:correctioncovariance},\ref{seq:correction_diffusion},\ref{seq:correctiongamma}) are in general dimensions, however only in \(d=1\) is the perturbative RG likely to yield good results. The flows close under the \(\mathbb{R}^{4}\) space \(\{T,\, X,\,Y,\,Z\}\): \(T\) is the ratio of diffusion coefficients, \(X\) and \(Y\) are non-linear cross-couplings, and \(Z\) is the KPZ effective non-linearity for the dynamical mode \(\tilde{\theta}_{1}\). The flows are:
\begin{equation}
\begin{aligned}
\partial_{\ell} T &= \frac{Z}{\pi} \left(-\frac{2 X ((T-3) X-3 T Y+Y)}{(T+1)^2}-T-X
   Y\right)\\\partial_{\ell} X &= -\frac{4 Z (X-1) X (X-Y)}{\pi  (T+1)^2}\\\partial_{\ell} Y &=\frac{Z Y (X-Y) \left(2 (5 T+1) X+(T+1)^2 Y-4 T (T+1)\right)}{\pi  T
   (T+1)^2}\\\partial_{\ell} Z &=Z \left(\frac{Z (Y (Y-3 X)-2 T)}{\pi  T}+1\right)
\end{aligned}
\label{seq:rg_flow_equations}
\end{equation}
where all flows feature a multiplicative factor of \(Z\). Solving these equations involves evolving the bare parameters at \(\ell=0\) (initial RG time) to \(\ell \rightarrow \infty\). The flows for \(\{X, Y, Z\}\) are self-generated, meaning \(\partial_{\ell}X = 0\) if X = 0. For bare parameters lying in Quadrants I and III, the flows predict a line of fixed points along \(X=Y\) as shown in FIG.~\ref{fig:combined}(iii) enforcing \(\chi=1/2\). The flows for all \(\Gamma\) vanish along the FDR and \(\chi=1/2\), giving rise to KPZ scaling \(z=3/2\) for both modes. For different initial points, the flows intersect the FDR line at different values of \(X\), determining the fixed points \(T^{*}\) and \(Z^{*}\), where the asterisk denotes the fixed point as \(\ell \rightarrow\infty\). 
\begin{figure}[h!]
    \centering
    \includegraphics[width=0.85\linewidth]{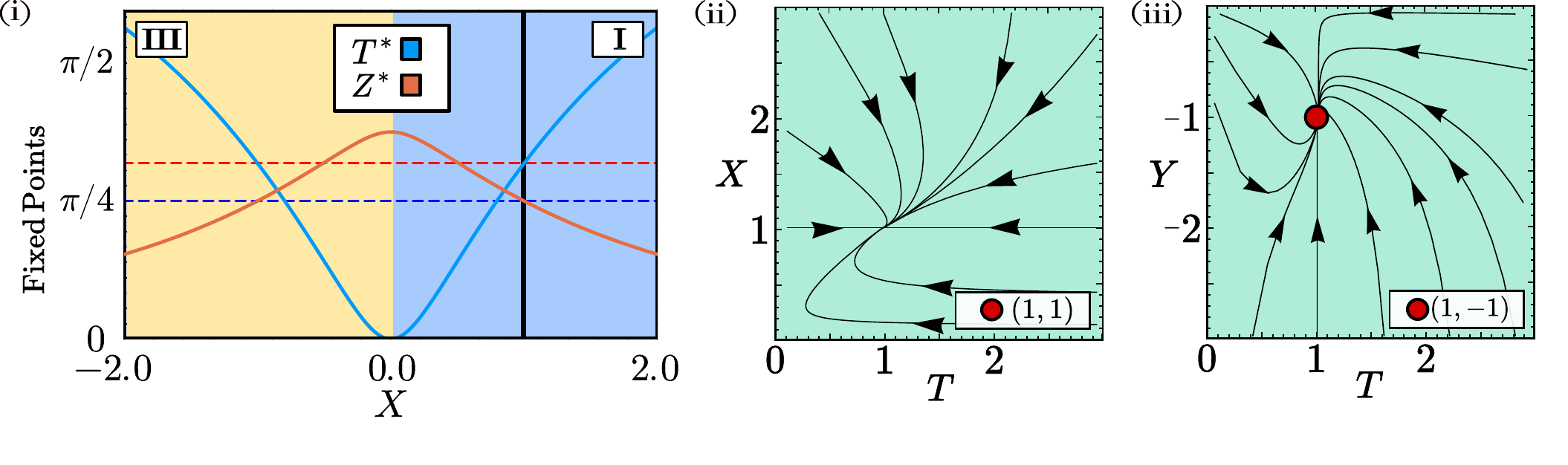}
    \caption{(i) Fixed points of the flows along the \(X=Y\) FDR line valid for Quadrants I and III. Along \(X=Y\), the intersection point determines the fixed point structure in the (T,Z) plane. The fixed points \(T^{*}\) and \(Z^{*}\) are shown in blue and red respectively. (ii-iii) Projected flows in Quadrant II in the \((T,X)\) plane at constant initial \(Y\) and in the \((T,Y)\) plane at constant initial \(X\), showing that the relevant attractive fixed point is \((X,Y,T)=(1,-1,1)\) (here \(Z\) diverges to infinity).}
    \label{fig:fixedpoints}
\end{figure}
Along the Cole-Hopf line when \(X=1\), the fixed points are \(Z^{*}=\pi/4\) and \(T^{*}=1\) supporting the decoupling hypothesis in agreement with the single-component case. There are two distinct behaviours along the Cole-Hopf line. When \(Y>0\), independent of the starting \(T\) we flow to a point where \(T^{*}=1\), \(X^{*}=Y^{*}=1\) where FDR is satisfied. Consequently, the renormalised parameters for the noise and interaction vertices satisfy \((\tilde{\Gamma}^{1*}_{22}\Delta_{22}^{*})/(\tilde{\Gamma}^{1*}_{11} \Delta_{11}^{*})=1\). At the fixed point there exists a transformation, \(\hat{\theta}^{\alpha}=s^{\alpha}_{\beta}\tilde{\theta}^{\beta}\) with
\begin{equation} s=\begin{pmatrix}\tilde{\Gamma}^{1*}_{11} & (\tilde{\Gamma}^{1*}_{11} \tilde{\Gamma}^{1*}_{22})^{1/2}\\ \tilde{\Gamma}^{1*}_{11}&-(\tilde{\Gamma}^{1*}_{11}\tilde{\Gamma}^{1*}_{22})^{1/2}
\end{pmatrix}
\label{eq:decouplingColeHopfsup}
\end{equation}
where the equations decouple symmetrically 
\begin{equation}
\begin{aligned}
     \partial_{t} \hat{\theta}_{1} &= \tilde{D}_{11}^{*}\partial_{x}^{2}\hat{\theta}_{1} +\dfrac{1}{2} (\partial_{x}\hat{\theta}_{1})^{2}  + \tilde{\Gamma}^{1*}_{11}\sqrt{2\tilde{\Delta}_{11}^{*}}\left(\zeta_{1}+\zeta_{2}\right)\\ 
     \partial_{t} \hat{\theta}_{2} &= \tilde{D}_{11}^{*}\partial_{x}^{2}\hat{\theta}_{2} +\dfrac{1}{2} (\partial_{x}\hat{\theta}_{2})^{2}  + \tilde{\Gamma}^{1*}_{11}\sqrt{2\tilde{\Delta}_{11}^{*}}\left(\zeta_{1}-\zeta_{2}\right).
\end{aligned}
\end{equation}
In the decoupled coordinates, there is a definite scaling hypothesis for the two-point correlation functions and coordinate transformations come from rotating KPZ modes into one another. Experimentally, measurements are made in the original \(\mathbb{Z}_{2}\) coordinates and so we need to use Eq.~(\ref{eq:decouplingColeHopfsup}) and the inverse of the normal mode transformation to get analytical forms in the physical coordinates.

In the unstable Quadrant II where \(Y<0\), there is a complex Cole-Hopf solution which can be motivated by starting with the complex stochastic heat equation \(\partial_{t} \Psi = \partial_{x}^{2}\Psi + \Psi \xi \) for \(\Psi \in \mathbb{C}\), substituting \(\tilde{\theta} = \log{\Psi}\), and expanding \(\tilde{\theta} = a\tilde{\theta}_{1}+i b \tilde{\theta}_{2}\). This maps onto the Cole-Hopf line with \(X=1\) and \(Y=-b^{2}/a^{2}\). We ignore this quadrant for stability reasons, however the RG predicts a stable fixed point in the subspace \((X,Y,T)=(1,-1,1)\), but \(Z\) diverges prompting the onset of large height fluctuations, breaking any scaling hypothesis. We highlight the unstable fixed point at \((X,Y,T,Z)=(1,0,1,\pi/2)\) on the boundary between Quadrant I and II which corresponds to having an independent KPZ mode which additionally acts as a multiplicative noise on the second mode.   

As \(X\rightarrow0\) along the FDR, the fixed point corresponds to \(T\rightarrow0\), leading to a numerical instability as the theory becomes noise-dominated. The parameter \(T\) controls the relative relaxation of the modes suppressing height fluctuations. In this \textit{inviscid} limit around small \(X,\,Y\), the expected long-time KPZ scaling behaviour is poor. At the unstable fixed point \(X*=Y*=0\), the equations in normal mode coordinates decouple with one KPZ mode and one EW mode. At intermediate times and lengthscales, the \(\mathbb{Z}_{2}\) basis rotates distinct scaling modes into one another. This behaviour is hinted at in the numerical simulation at \(X=1/2, \, Y=1/2\) in Section D where the fluctuation distribution for \(\tilde{\theta}_{2}\) is more Gaussian with smaller kurtosis, suggesting the need for better methods to probe this extremal point. The flows also predict a similar \(T\rightarrow0\) for bare parameters originating in Quadrant IV but we discount it based on non-hyperbolicity.
\section{D. Direct Simulations of Quadrant I}
\label{sec:supp-numerics}
The analytical discussion relies on the accurate description of the scaling by perturbative RG around the decoupled point \((X,Y)=(1,1)\). However, it does not rule out the possibility of non-perturbative effects altering the scalings. To provide a more comprehensive understanding, we perform direct simulations of the coupled KPZ equations in Quadrant I. We employ the forward Euler-–Maruyama method to directly integrate the stochastic fields \(\theta_{i,t_{i}}\). It is well-known that the finite difference discretisation is crucial in ensuring correct behaviour, especially in preventing instabilities from large height differences between neighbouring lattice points \(\Delta h > h_{c} \sim O\left({\Delta\Gamma^{2}}/{D^{3}}\right)^{-1/2}\). The discretised coupled KPZ equation can be written as a system of stochastic ODEs, with two component degrees of freedom \(\alpha\) at each lattice point
\begin{equation}
    \theta^{\alpha}_{t_{i+1}} = \theta^{\alpha}_{t_{i}} + \left[D^{\alpha}_{\beta} L^{\beta}[\theta] + \Gamma^{\alpha}_{\beta\gamma} N^{\beta}[\theta]N^{\gamma}[\theta]\right]\Delta t + \Delta W^{\alpha}_{t_{i}}.
\end{equation}
where \(L[\theta_{i}^{\alpha}]=(\theta^{\alpha}_{t_{i},x+1}-2\theta^{\alpha}_{t_{i},x}+\theta^{\alpha}_{t_{i},x-1})/2\Delta x\) is the discretised Laplacian operator. We set \(\Delta t = 0.01\) to ensure stability. The fields are integrated from flat initial conditions \(\theta^{\alpha}(t=0)=0\) to \(t=800\). The noise term \(\Delta W_{t_{i}}\) follows a centred Gaussian distribution with increments \(\Delta W^{\alpha}_{t_{i}} = W^{\alpha}_{t_{i}} - W^{\alpha}_{t_{i-1}} \sim N(0, 2\tilde{\Delta}_{\alpha\alpha} \Delta t) \), which reproduce the covariance matrix. The white noise process is expressed as the sum of two independent processes \(\Delta W^{\alpha}_{t_{i}} =  \sqrt{12\Delta t} R_{1,t_{i}} \sqrt{2\tilde{\sigma}_{1}} + \sqrt{12\Delta t} R_{2,t_{i}} \sqrt{2\tilde{\sigma}_{2}} \) where \(\tilde{\sigma}_{i} = {(\Delta_{11}\pm \sqrt{\Delta_{11}^{2}-\Delta_{12}^{2}})}/{2}\) and is approximated from sampling a centered uniform distribution \(R_{i,t_{i}} \sim U([-0.5,0.5])\). The noise discretisation is common in numerics of stochastic equations as a Wiener process can be generated as a limiting process of a random walk with equivalent first and second moments. For the non-linear operator \(N\), we define:
\begin{equation}
    N^{\alpha}[\theta_{j}] = \left(\dfrac{\theta^{\alpha}_{j+1}-\theta^{\alpha}_{j-1}}{2\Delta x}\right).
\end{equation}
We set all \(D_{11}=\Delta_{11}=1\) and \(\Gamma^{1}_{11}=-1\), thus \(Z_{0}\sim 1\) which limits instabilities in the height surface while still showcasing KPZ scaling. There were two primary goals: (a) to characterise the exponents for various sizes of system, verifying convergence to \(\chi=1/2\) and \(z=3/2\) for both dynamical modes in rotated and unrotated coordinates even away from the \(X=Y\) line, and (b) to characterise the distribution of fluctuations on the FDR manifold X=Y and the Cole-Hopf line. This numerical investigation allows us to test and verify the scaling predictions from the RG analysis, providing insights into both the perturbative and non-perturbative behavior of the coupled KPZ system.
\subsection{Dynamical Exponents}
To classify the dynamical exponents in Quadrant I, we investigate how the scaling behaves in the long system size and time limit. The length of the system is varied across \(\{2^{8}, 2^{12}, 2^{16}, 2^{20}\}\), and we analyse the finite size convergence of the roughness \(\alpha\), growth  \(\beta\), and dynamical \(z\) exponents. We consider two points I.A. and I.B., shown in TABLE~\ref{table:betas}, located away from the FDR line \(X=Y\) and Cole-Hopf line \(X=1\). In the single-component theory, the surface roughness 
\begin{equation}
     W_{\alpha}(t) = \langle \overline{(\theta_{\alpha})^{2}} - \overline{\theta_{\alpha}}^{2} \rangle  \sim t^{2\beta}, \; \; \; t \ll L^{z} 
\end{equation}
can be fitted to extract the growth exponent \(\beta\), demonstrating a crossover from EW \(\beta=1/4\) to KPZ \(\beta=1/3\) in the long wavelength limit. 
\begin{figure}[h!]
    \centering
        \centering
        \includegraphics[width=0.75\linewidth]{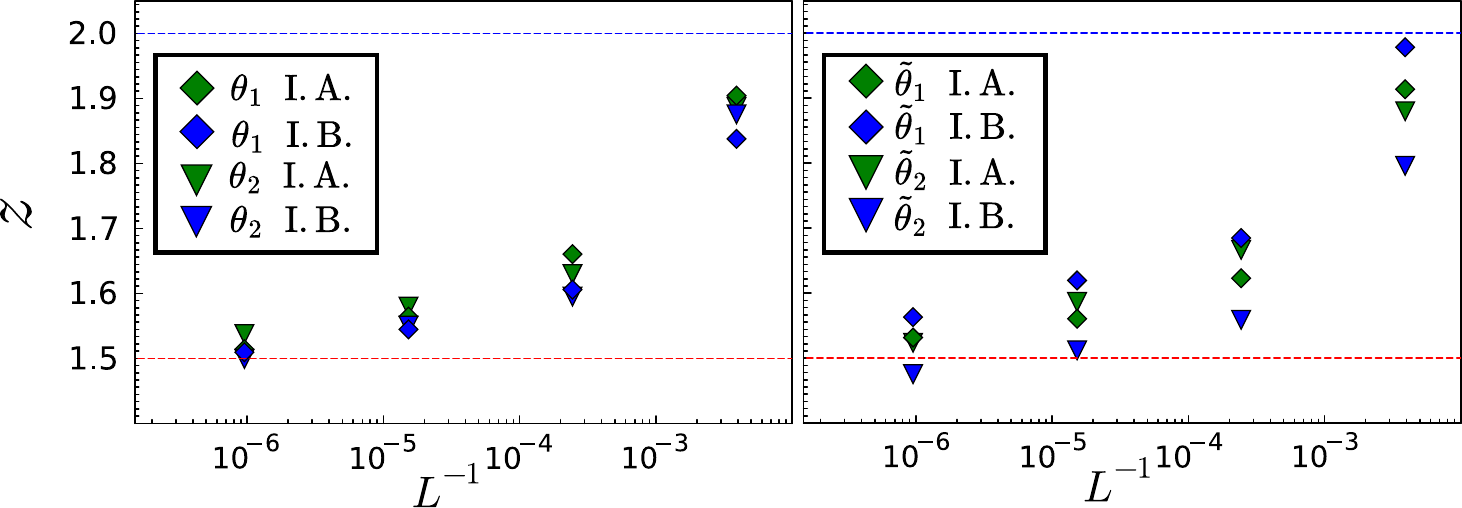}
    \caption{The dynamical exponent \(z\) plotted against \(1/L\) where Subfigure (i) shows the finite scaling analysis for the unrotated \(\mathbb{Z}_{2}\) coordinates and Subfigure (ii) shows convergence in the normal mode basis. In both cases we get the expected crossover behaviour from \(z=2\) EW behaviour to \(z=3/2\) KPZ scaling.}
\label{seq:convergence_dynamical_exponent}
\end{figure}
While simulations tend to underestimate \(\beta\) in the growth phase, the values are within expected error margins. This underestimation can be attributed to the XY to KPZ crossover. The scaling of the two-point same-time correlation function 
\begin{equation}
    \hat{C}_{\alpha\beta}(|x-x'|) = \langle \left(\theta_{\alpha}(x+x')-\theta_{\beta}(x) \right)^{2}\rangle  = C_{\alpha\beta} |x-x'|^{2\chi}.
\end{equation}
can be fitted to extract the roughness exponent \(\alpha\) from which the dynamical exponent is fixed by their ratio. The fitted values for \(L=2^{20}\) are plotted in FIG.~\ref{seq:convergence_dynamical_exponent}. The full list of initial values and fitted values is contained in TABLE~\ref{table:betas}. For points initialised on the FDR line, the correlation functions satisfy \(\tilde{C}_{11}=\tilde{C}_{22}\approx 1\), which aligns with the same-time correlations remaining their bare values under RG strictly imposed by emergent Ward-Takahashi identities. This verifies that the FDR line is stationary under RG. Away from the FDR line, such as at points I.A. and I.B., the correlations flow to a non-unity value, suggesting that the correlation functions are renormalised as shown in TABLE~\ref{table:betas}. 
\begin{table}[htbp]
    \centering
    \begin{tabular}{||@{}l||c|c|c|c||c|c|c|c|c||c|c|c|c|c||c|c|c|c|c||@{}} 
        \toprule
        \hline
        \multirow{2}{*}{} & \multicolumn{4}{c|}{\textbf{RG I.C.}} & \multicolumn{5}{c|}{\textbf{Unrotated \(\theta_{1}\)}} & \multicolumn{5}{c|}{\textbf{Rotated \(\tilde{\theta}_{1}\)}}& \multicolumn{5}{c|}{\textbf{Rotated \(\tilde{\theta}_{2}\)}}  \\ \hline
        & \textbf{$X$} & \textbf{$Y$} &  \textbf{$T$} & \textbf{$W$} & \textbf{$v_{\infty}$} & \textbf{$\tilde{M}_{\infty}$} & \textbf{$\beta$} & \textbf{$\alpha$} & \(C_{11}\)& \textbf{$v_{\infty}$} & \textbf{$\tilde{M}_{\infty}$} & \textbf{$\beta$} & \textbf{$\alpha$} & \(\tilde{C}_{11}\)& \textbf{$v_{\infty}$} & \textbf{$\tilde{M}_{\infty}$} & \textbf{$\beta$} & \textbf{$\alpha$} & \(\tilde{C}_{22}\)\\ \hline
        \midrule
        \textbf{CH} & \(1.0\) & \(2.0\) & \(0.5\) & \(2.0\) & -0.200 & 0.546 & 0.316 & 0.484 & 1.048 & -0.283 & 0.449 & 0.281 & 0.467 & 1.026 & 0.000 & 0.649 & 0.281 & 0.498 & 1.073\\ \hline
        \textbf{FDR} & \(1.0\) & \(1.0\) & \(1.0\) & \(1.0\) & -0.248 & 0.512 & 0.298 & 0.466 & 1.042 & -0.35 & 0.512 & 0.298 & 0.466 & 1.042 & 0.000 & 0.512 & 0.298 & 0.466 & 1.042\\ \hline
        \textbf{FDR} & \(2.0\) & \(2.0\) & \(1.0\) & \(1.0\) & -0.212 & 0.518 & 0.298 & 0.467 & 1.040 & -0.30 & 0.469 & 0.29 & 0.465 & 1.046 & 0.000 & 0.569 & 0.290 & 0.468 & 1.035\\ \hline
        \textbf{FDR} & \(0.5\) & \(0.5\) & \(1.0\) & \(1.0\) & -0.297 & 0.502 & 0.306 & 0.476 & 1.027  & -0.42 & 0.609 & 0.319 & 0.478 & 1.02 & 0.000 & 0.403 & 0.319 & 0.475 & 1.034\\ \hline
        \textbf{I.A.} & \(1.07 \)& \(0.73\) & \(1.11\) & \(0.82\) & -0.242 & 0.51 & 0.306 & 0.472 & 1.041 & -0.342 & 0.545 & 0.295 & 0.464 & 1.195 & 0.000 & 0.475 & 0.295 & 0.483 & 0.889\\ \hline
        \textbf{I.B.} & \(0.98 \)& \(2.18\) & \(1.63\) & \(0.75\) & -0.324 & 0.582 & 0.309 & 0.477 & 1.102 & -0.458 & 0.882 & 0.309 & 0.474 & 1.483 & -0.001 & 0.327 & 0.309 & 0.484 & 0.721\\
        \bottomrule
    \end{tabular}
    \caption{Scalings for different initial bare parameters in Quadrant I. \(v_{\infty}\) and \(\tilde{M}_{\infty}\) refer to the fitted values at \(t=\infty\) for \(L=2^{20}\). Static properties such as \(C_{\alpha\beta}\) are calculated at time shot \(t=800\).}
    \label{table:betas}
\end{table}
\subsection{Distributions of Fluctuations}
The distributions of fluctuations along the Cole-Hopf line highlight that the RG flow decouples the KPZ equations, leading to fluctuations that are the sum and difference of TW-GOE distributions. This is evident from the simulated distributions agreeing with the expected distributions in yellow, as well as the moments approaching their expected values over long times. For \((X,Y)=(1/2,1/2)\), the rotated coordinates \(\tilde{\theta}_{2}\) have less kurtosis, approaching a Gaussian. This smooth crossover along the FDR line from TW-GOE to Gaussian is expected, as at \(X=0\) the equations are decoupled with a KPZ mode and an EW mode. This behaviour should be apparent at finite time and finite system-size, however whether it remains in the long-time limit is not explored. At point \((X,Y)=(2,2)\), \(\tilde{\theta}_{1}\) displays a longer tail to the right indicating the reverse effect. These results clearly showcase the crossover effect as the system moves across the FDR line, with distinct fluctuation behaviors emerging in the different regions.
\begin{figure}[h!]
    \centering
        \centering
        \includegraphics[width=\linewidth]{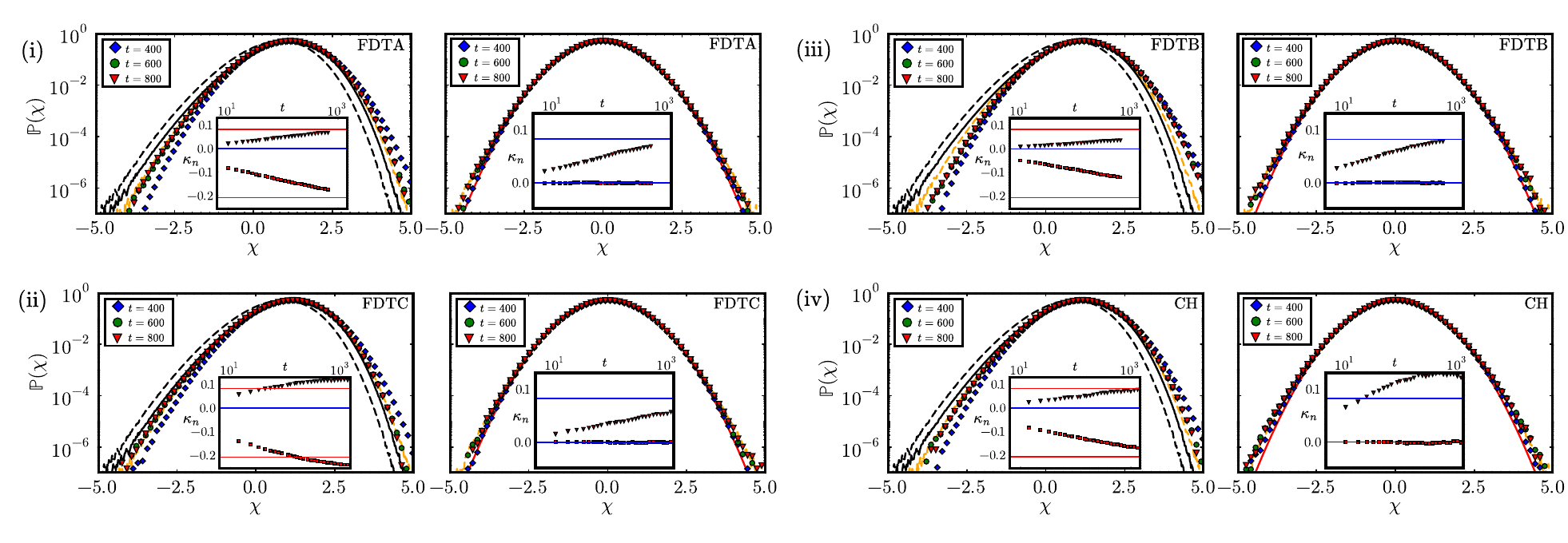}
    \caption{ Distribution of fluctuations for \(\tilde{\theta}_{1}\) and \(\tilde{\theta}_{2}\) on the left and right respectively for different initial bare parameter values in Quadrant I. The subfigures within each graph show the convergence of the moments of the distributions which are the skewness and kurtosis. (i) FDTA corresponds to \((X,Y)=(1,1)\) at the decoupled point. (ii) FDTB corresponds to \((X,Y)=(2,2)\) (iii) FDTC corresponds to \((X,Y)=(1/2,1/2)\), (iv) CH correspondings \((X,Y)=(1,2)\). All parameters and relevant scalings can be found in Table \ref{table:betas}.}
\label{seq:fluctuations_sm}
\end{figure}
\section{E. Spacetime Vortex Dominated Phase}
As the intercomponent interaction \(v_{c}\) is tuned from the decoupled point \((X,Y)=(1,1)\) into Quadrant IV where the theory is non-hyperbolic, the coupled KPZ equations exhibit an onset of large spatial fluctuations \(\theta(x_{i},t_{j})-\theta(x_{i+1},t_{j}) > 2\pi\) in the unwrapped phase. Physically, the phase is a compact variable and the large height fluctuations give rise to a spacetime vortex (STV) dominated phase. 
\begin{figure}[h!]
    \centering
        \centering
        \includegraphics[width=0.6\linewidth]{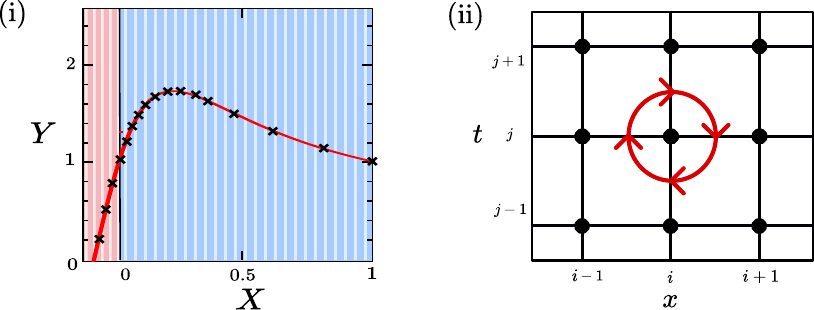}
    \caption{(i) Tuning \(\tilde{v}_{c}\) leaving all other parameters constant traces out a line in the \((X,Y)\) plane which allows you to tune the theory from Quadrant I (blue) where we expect KPZ to Quadrant IV where we expect the onset of an STV phase. (ii) Lattice point of the unravelled phase in time and definition of the spacetime vortex as a non-zero circulation around a point on a lattice.}
\label{seq:sm_vortices}
\end{figure}
This instability arises from the balance between the diffusion constant, acting as viscosity suppressing fluctuations, and the non-linear term driving the system into a rough geometry. The KPZ theory describes this rough geometry, and the crossover rate from flat to rough geometry is related to the size of the restorative diffusion term. A spacetime vortex is a point on the lattice where there is non-zero circulation as shown in FIG.~\ref{seq:sm_vortices}. The charge of the vortex \(c^{*}\) is
\begin{equation}
    c^{*} = \dfrac{1}{2\pi}\left[\mathcal{U}(\theta_{i,j+1},\theta_{i+1,j}) + \mathcal{U}(\theta_{i-1,j},\theta_{i,j+1})+ \mathcal{U}(\theta_{i,j-1},\theta_{i-1,j})+ \mathcal{U}(\theta_{i+1,j},\theta_{i,j-1}) \right]
\end{equation}
where \(\mathcal{U}\) calculates the unravelled phase difference between two points. In systems with \(U(1)\times U(1)\) symmetry, the elementary topological excitations are half-vortices in either component in the unrotated \(\mathbb{Z}_{2}\) theory. 

We simulate the SCGLE (\ref{eq:coupled_SCGLE}) in the low noise regime with \(\gamma_{T}=0.1\) and time step \(dt=0.01\) up to \(t_{\mathrm{tot}}=10^{5}\) using a system size \(L_{x}=2048\). The total time is much larger than the saturation time \(L_{\mathrm{sat}}\) which is proportional to \(L_{x}^{z}\) where \(z\) is the KPZ dynamical exponent. Tuning \(\tilde{v}_{c}\) defines a line \(\gamma:[-1.2,0] \rightarrow \mathbb{R}^{2}\) for \(\tilde{v}_{c}\mapsto (X,Y)\) as shown in FIG.~\ref{seq:sm_vortices}(i). Parameters are set \(v_{d}=0\), \(k_{d}=u_{d}=\mu_{d}=1\) and \(k_{c}=3,\;u_{c}=1.5\). \(\mu_{c}\) is determined by solving the consistency relation \(\mu_{c}= - (u_{c}+v_{c})\) with the density fixed \(\rho\) = 1. The chosen parameters mean that time \(t\) and space \(x\) are in units of \(\mu_{d}^{-1}\) and \(\sqrt{k_{d}}\) respectively. Setting \(v_{d}=0\) the mapping to \((X,Y)\) is: 
\begin{equation}
\begin{aligned}
    X &= \frac{2 (\mathcal{R}(k)-\mathcal{R}(u))}{\mathcal{R}(k) \tilde{v}_{c}+\mathcal{R}(k)-\mathcal{R}(u)}-1\\ 
    Y &=\frac{\left(\mathcal{R}(u)^2+(\tilde{v}_{c}-1)^2\right) (\mathcal{R}(k) \mathcal{R}(u)+\tilde{v}_{c}+1)}{\left(\mathcal{R}(u)^2+(\tilde{v}_{c}+1)^2\right) (\mathcal{R}(k) \mathcal{R}(u)-\tilde{v}_{c}+1)}
\end{aligned}
\end{equation}
which converges to \((1,1)\) in the decoupled limit \(\tilde{v}_{c}\rightarrow 0\). The time-unravelled phase is analysed across \(L_{t}=2000\) time sampling points. Consequently, the time spacing on the spacetime lattice is \(t_{\mathrm{tot}}/L_{t}=50\). The spatial unit is unity \(\Delta x = 1\). We have two measures for the onset of the STV phase, the probability of vortices on the spacetime lattice and the autocorrelation function. We define the autocorrelation function 
\begin{equation}
    \Delta_{\alpha}(t_{1}-t_{2}) = \overline{\langle (\theta_{\alpha}(t_{1})-\theta_{\alpha}(t_{2}))^{2}\rangle} - \overline{\langle (\theta_{\alpha}(t_{1})-\theta_{\alpha}(t_{2}))\rangle}^{2} \sim |t_{1}-t_{2}|^{2\beta}.
\end{equation}
as time-time correlations of locally defined points on the lattice. \(\langle \cdot \rangle\) indicates averaging over trajectories and \(\overline{\,\cdot\,}\) as a spatial average over lattice points. To ensure that we are probing long-time dynamics, we evolve the simulation to \(t_{2}=50000\) in the saturation regime \(L_{x}^{z}\), thus the autocorrelation function in FIG.~\ref{fig:figurestv}(i-ii) shows the systems between \(t=5\times 10^{4}\) to \(t=10^{5}\). For KPZ systems, starting from an flat initial conditions, we expect to see a crossover from EW scaling with \(\beta=1/4\) to \(\beta=1/3\) for large system sizes and large times. However, in the STV dominated phase, the growth exponent is \(\beta=1/2\) indicating a departure from the expected scaling. The other measure is the vortex probability \(\mathcal{P}_{v}\) which is defined as the ratio of the number of vortices in each component \(N_{v}\) to the total number of spacetime lattice points \(N_{v_{i}}/(L_{t}\times L_{x})\). The section of the spacetime lattice in FIG.~\ref{fig:figurestv}(iv) is from taking a spatial slice of \(100\) points and the final \(100\) time points on the lattice. We use periodic boundary conditions and expect that the solution is translationally invariant. The probability of a vortex across the whole spacetime lattice features a smooth crossover from a suppressed vortex phase in Quadrant I to a vortex dominated phase in Quadrant IV. The smooth nature of transition does not suggest that it is first-order. This provides unequivocal evidence that Quadrant IV is disordered and also poses an opportunity for experimentalists to analyse the onset of STV in physical systems.